\documentclass[journal]{vgtc}               

\usepackage{microtype}                
\PassOptionsToPackage{warn}{textcomp}  
\usepackage{textcomp}                  
\usepackage{amsmath}
\usepackage{graphicx}
\usepackage[]{algorithm2e}
\usepackage{mathtools}
\usepackage{amssymb}
\usepackage{threeparttable}
\usepackage{caption}
\usepackage{times}
         
\usepackage{cite}                      
\usepackage{tabu}                      
\usepackage{booktabs}                  

\usepackage{adjustbox} 
\usepackage{xcolor}

\onlineid{1092}

\vgtccategory{Research}

\vgtcpapertype{Algorithm / Technique}

\title{DeepOrganNet: On-the-Fly Reconstruction and Visualization of 3D / 4D Lung Models from Single-View Projections by Deep Deformation Network}

\author{Yifan Wang, Zichun Zhong, and Jing Hua}
\authorfooter{
\item
Yifan Wang, Zichun Zhong, and Jing Hua are with the Department of Computer Science, Wayne State University, Detroit, MI 48202. E-mail: \{yifan.wang2,zichunzhong,jinghua\}@wayne.edu.
\item
Corresponding author is Zichun Zhong.
}

\shortauthortitle{Biv \MakeLowercase{\textit{et al.}}: Global Illumination for Fun and Profit}

\abstract{This paper introduces a deep neural network based method, i.e., DeepOrganNet, to generate and visualize fully high-fidelity 3D / 4D organ geometric models from single-view medical images with complicated background in real time. Traditional 3D / 4D medical image reconstruction requires near hundreds of projections, which cost insufferable computational time and deliver undesirable high imaging / radiation dose to human subjects. Moreover, it always needs further notorious processes to segment or extract the accurate 3D organ models subsequently. The computational time and imaging dose can be reduced by decreasing the number of projections, but the reconstructed image quality is degraded accordingly. To our knowledge, there is no method directly and explicitly reconstructing multiple 3D organ meshes from a single 2D medical grayscale image on the fly. Given single-view 2D medical images, e.g., 3D / 4D-CT projections or X-ray images, our end-to-end DeepOrganNet framework can efficiently and effectively reconstruct 3D / 4D lung models with a variety of geometric shapes by learning the smooth deformation fields from multiple templates based on a trivariate tensor-product deformation technique, leveraging an informative latent descriptor extracted from input 2D images. The proposed method can guarantee to generate high-quality and high-fidelity manifold meshes for 3D / 4D lung models; while, all current deep learning based approaches on the shape reconstruction from a single image cannot. The major contributions of this work are to accurately reconstruct the 3D organ shapes from 2D single-view projection, significantly improve the procedure time to allow on-the-fly visualization, and dramatically reduce the imaging dose for human subjects. Experimental results are evaluated and compared with the traditional reconstruction method and the state-of-the-art in deep learning, by using extensive 3D and 4D examples, including both synthetic phantom and real patient datasets. The efficiency of the proposed method shows that it only needs several milliseconds to generate organ meshes with 10K vertices, which has a great potential to be used in real-time image guided radiation therapy (IGRT).}


\CCScatlist{ 
 \CCScat{K.6.1}{Management of Computing and Information Systems}
{Project and People Management}{Life Cycle};
 \CCScat{K.7.m}{The Computing Profession}{Miscellaneous}{Ethics}
}

\vgtcinsertpkg

\begin{document}

\firstsection{Introduction}

\maketitle

Cone beam computed tomography (CBCT) has become increasingly important in cancer radiotherapy for understanding the anatomical structure of organs and pinpointing tumors during the treatments. Integrated CBCT is an important and convenient tool for patient positioning, verification, and visualization in image guided radiation therapy (IGRT). Traditional high-quality CBCT image reconstruction requires near hundreds of projections, which consequently deliver undesired high imaging / radiation dose to patients as well. The high imaging dose to healthy organs in CBCT scans~\cite{islam2006patient,kan2008radiation,song2008dose} is a crucial clinical concern. Practically, the imaging dose in CBCT can be reduced by reducing the number of X-ray projections and lowering tube voltage setting. On the other hand, due to the limited number of projections, the image quality is highly degraded in 3D-CBCT reconstructed by conventional methods, such as Feldkamp-Davis-Kress (FDK)~\cite{feldkamp1984practical} (plenty of artifacts and noises with lower accuracy). Several strategies have been proposed to enhance the image quality of reconstructed CBCT. One major kind of approaches is to use iterative image reconstruction algorithms, such as simultaneous algebraic reconstruction technique (SART)~\cite{andersen1984simultaneous}, total variation (TV) minimization~\cite{song2007sparseness}, and prior image constraint techniques~\cite{chen2008prior,brock2010reconstruction}. Thus, the accuracy of subsequent 3D organ modeling does highly depend on the quality of the reconstructed images. Currently, the doctors and clinicians need to use some post-processing methods / tools to segment, reconstruct, and visualize the 3D organ models, which is quite time-consuming and cumbersome.

Recently, there is an emerging trend to generate 3D models by deep neural network in computer vision, computer graphics, and visualization communities, in which the 3D shapes can be captured and represented from the input raw data of different formats such as 3D point clouds, 3D volumes, multi-view 2D images, etc. Among them, deriving the 3D shape from a single view is fundamental and very challenging. Recently, deep learning techniques have been developed to generate / reconstruct 3D shapes from a single RGB natural image (e.g., photograph)~\cite{choy20163d,fan2017point,kurenkov2018deformnet,wang2018pixel2mesh}. Their 3D shape outputs from the neural network can be represented in different formats, such as a volume~\cite{choy20163d}, point loud~\cite{fan2017point,kurenkov2018deformnet}, or surface mesh~\cite{wang2018pixel2mesh}. However, these methods either require complicated post-processing to generate the surface models~\cite{choy20163d,fan2017point}, or have non-manifold and invalid surface elements~\cite{wang2018pixel2mesh}.

In order to build the bridge to directly generate the 3D shape meshes from a single 2D medical grayscale image on the fly, in this work, we present a deep neural network based method, i.e., \emph{DeepOrganNet}, to generate and visualize high-fidelity fully 3D / 4D organ geometric models from single-view medical images, e.g., 3D / 4D-CBCT projections, by learning the smooth deformation fields based on a trivariate tensor-product deformation technique. Experimental results are evaluated and compared with the traditional reconstruction method and the state-of-the-art in deep learning, by using extensive 3D and 4D examples, including both synthetic phantom and real patient datasets. The key \emph{contributions} of our work are as follows:\vspace{-2mm}
\begin{itemize}
 \item It proposes an end-to-end deep learning method with a lightweight and effective neural network to reconstruct multiple high-fidelity 3D organ meshes with a variety of geometric shapes from a single-view medical image with complicated background and noises.\vspace{-2mm}
 \item The proposed organ reconstruction network simultaneously learns the optimal selection and the best smooth deformation from multiple templates via a trivariate tensor-product deformation technique, i.e., free-form deformation (FFD), to match the query 2D image.\vspace{-2mm}
 \item To our knowledge, it is the first time using deep learning framework to generate multiple 3D organ meshes (such as left and right lungs in our application) from a single-view medical image.\vspace{-2mm}
  \item The application and user study on IGRT demonstrate that the accurate on-the-fly tracking and reconstruction of 3D / 4D organ shapes facilitated by our method have the potential in improving the current IGRT procedure and practice.\vspace{-2mm}
\end{itemize}

\section{Related Work}
In this section, we only review some most related work on 3D shape reconstruction from single images in computer vision / graphics, visualization, and medical imaging domains.

\subsection{3D Shape from Single-View Image in Computer Vision}
In computer vision, graphics, and visualization, 3D reconstruction is the process of capturing the shape and appearance of real objects.
\subsubsection{Traditional Learning Based Methods}
Hoiem et al.~\cite{hoiem2005automatic} and Saxena et al.~\cite{saxena2009make3d} started to use statistic and learning based approaches for 3D shape reconstruction from a single image several decades ago. Recently, Kar et al.~\cite{kar2015category} proposed to learn category-specific 3D shape models from object silhouettes and then capture intra-class shape variation from a single image. Carreira et al.~\cite{carreira2016lifting} proposed a method to estimate the camera viewpoint using rigid structure-from-motion and then reconstruct object shapes by optimizing over visual hull proposals guided by loose within-class shape similarity assumptions. Fouhey et al.~\cite{fouhey2013data} demonstrated to learn their proposed primitives to infer 3D surface normals given a single image. Eigen et al.~\cite{eigen2014depth} presented a method to estimate and find 3D depth relations from a single stereo image by using a multi-scale deep network.

With the help of ShapeNet~\cite{chang2015shapenet}, a richly-annotated and large-scale repository of 3D CAD models, there are several 3D reconstruction approaches presented in the recent few years. For instances, Huang at al.~\cite{huang2015single} proposed a joint analysis method for shape reconstruction by estimating the camera pose, computing dense pixel-level correspondences between image patches, and finally creating a 3D model for each image by an optimization.

\subsubsection{Deep Learning Based Methods}
\label{sec:dl_methods}
Most recently, using deep learning methods to analyze and represent 3D objects is becoming a popular trend, inspired by the successes of these techniques in 2D images and 1D texts. Choy et al.~\cite{choy20163d} proposed a 3D recurrent reconstruction neural network (3D-R2N2) to output a reconstruction of the object with a 3D occupancy grid format, which cannot well preserve the surface geometry of a 3D shape. In order to predict a nicer surface space, Fan et al.~\cite{fan2017point} explored the generative networks for 3D geometry based on a point cloud representation. Kuryenkov et al.~\cite{kurenkov2018deformnet} proposed a DeformNet to achieve smooth geometric deformations in point clouds for 3D shape reconstruction. However, it is well-known that a 3D point cloud may not be as efficient and effective in representing the underlying continuous 3D geometry as compared to a 3D surface mesh. It needs some non-trivial post-processings to generate the valid surface meshes (to guarantee the manifold property).

Wang et al.~\cite{wang2018pixel2mesh} adopted a graph-based convolutional neural network to produce the 3D geometry by progressively deforming an ellipsoid with leveraging perceptual features extracted from an input image. However, this method can only reconstruct a single genus-0 topology shape, since the initial shapes are all deformed from an ellipsoid. Another limitation is that their deformation is defined on the surface space with a linear transformation model, which is difficult for the network to compute high-fidelity large deformation to accurately capture the shape geometry and they need several regularization terms to control the shape smoothness and local consistency. Smith et al.~\cite{Smith2019} extended Wang et al.'s work~\cite{wang2018pixel2mesh} by using an adaptive face splitting strategy in order to better capture the local surface geometry, but it still has the problems of having non-manifold elements and topological constraint (by using a sphere initial shape). The above methods are based on \emph{surface deformation}. However, one of the major limitations of surface deformation, whose deformation field is directly defined on the shape surface, is that its computational effort and numerical robustness are highly related to the complexity and quality of the surface tessellation~\cite{botsch2010polygon}. In the presence of the degenerate or poor quality triangles, the local transformations on these triangles are not well defined and thus lead to topological or non-manifold errors~\cite{wang2018pixel2mesh}, as shown in Sec.~\ref{sec:experiments}. Even with quite some efforts for adding regularization terms, such as Laplacian regularization, edge length regularization, etc.~\cite{wang2018pixel2mesh,Smith2019}, it is still difficult to fully guarantee the deformation consistency in the local vertex neighborhood.

From the mathematical aspect, this problem can be avoided by \emph{space deformation}. The key idea is to deform the ambient space (i.e., 3D volume space) enclosing the shapes, and thus implicitly deform the embedded surface shape (i.e., 2-manifold)~\cite{botsch2010polygon}. Compared with the surface-based deformation methods, space deformation approaches apply a trivariate deformation function to transform all the points of the original surface. One major advantage of the space deformation is that it does not depend on any particular surface representation, so that it can be used to deform all kinds of explicit surface representations, such as vertices of meshes or samples of point clouds~\cite{botsch2010polygon}. Classical free-form deformation (FFD)~\cite{sederberg1986free} represents the space deformation by a trivariate tensor-product spline function. Ponte et al. \cite{pontes2018image2mesh} proposed a learning framework (i.e., Image2Mesh) to reconstruct the single object 3D mesh from a 2D natural image by first deforming a selected template using symmetric FFDs and then linearly combining a few more strongly related templates. However, their method predominantly relies on a complicated and pre-computed graph embedding of templates and their framework is not end-to-end trainable. Jack et al.~\cite{jack2018learning} proposed a method to learn FFDs for multiple templates to infer a 3D shape reconstruction from a single natural image with a plain background, but their framework is also limited to generate a single object, without considering multi-object scenario.

Besides that, all the aforementioned deep learning based methods for 3D shape reconstruction from a single image are not designed and applied to medical image reconstruction and visualization.

\subsection{3D Volume (Shape) from Single-View Image in Medicine}
In medical image, 3D reconstruction is the process of computing the structure and tissue of real objects (not only the shape).
\subsubsection{Volumetric Image Reconstruction Methods}
Li et al.~\cite{li2010single,li2010real} utilized a deformable image registration method to compute deformation vector fields (DVFs) for the reference of a lung motion model. Then, a principal component analysis (PCA) based lung motion model has been applied to generate a motion vector field so as to reconstruct a volumetric image and locate 3D tumor from a single CBCT / X-ray projection. The algorithm was implemented on graphics processing unit (GPU) to achieve real-time efficiency. However, there are limitations of their method, such as a linear relationship between the image intensity of the computed and measured projection images may not be accurate. Some pre-processings for DVF computation are needed. The framework settings are not fully automatical and practical for clinical use. The single-view reconstruction suffers from an ill-posed problem because only one angle data is used in the reconstruction. To alleviate this issue, Liu et al.~\cite{liu2014wavelet} tried a wavelet-based reconstruction approach to the acquired singe-view measurements, but the reconstruction quality is still not satisfactory for clinical applications. Recently, Henzler et al. \cite{henzler2018single} proposed a convolutional encoder-decoder network to reconstruct a 3D volume from a 2D single-view cranial X-ray image. The direct coarse output is then improved to higher resolution by post fusion. The resulting 3D structure is still embedded in a 3D volume and the 3D shapes can only be shown by volume rendering with the manual-setting isosurface threshold.

\subsubsection{Shape Reconstruction Methods}
There are few works on directly reconstructing the 3D shapes (meshes) from medical images (grayscale pixels), since it is a cross-modality problem, which is relatively challenging. The traditional solution is to reconstruct the 3D volumetric images from multiple 2D view images at first~\cite{feldkamp1984practical,andersen1984simultaneous,song2007sparseness,chen2008prior,brock2010reconstruction}, and then use image segmentation methods to extract the region of interest (ROI), such as organs or tumors; and finally generate the 3D shape meshes (i.e., isosurface) by using Marching Cubes algorithm~\cite{lorensen1987marching}. For instance, iso2mesh~\cite{fang2009tetrahedral} is an open-source toolbox for generating 3D surficial and volumetric meshes from binary and grayscale images, but it needs to undergo the tedious procedure due to the complicated substeps.

Some researchers investigated to fill the gap to directly build the 3D shape from a limited / sparse number of 2D medical images. Fleute et al.~\cite{fleute1999nonrigid} proposed to use a few X-ray images generated from a C-Arm and to build the 3D shape of the patient bones or organs by deforming a statistical 3D model to the contours segmented on the X-ray views. Tang et al.~\cite{tang20052d} used a hybrid 3D atlas shape model to reconstruct or deformably register the surface of an object from two to four 2D X-ray projections of the object. Lamecker et al.~\cite{lamecker2006atlas} presented a method to reconstruct 3D shapes from few digital X-ray images on the basis of 3D-statistical shape models; however, there are some empirical pre-processings needed, such as thickness of the shape model, silhouette extraction, etc. Sadowsky et al.~\cite{sadowsky2006projected} presented a method for volume rendering of unstructured grids, which was applied in visualizing ``2D--3D'' deformable registration of anatomical models. Ehlke et al.~\cite{ehlke2013fast} proposed a novel GPU-based approach to render virtual X-ray projections of deformable tetrahedral meshes, and applied the method to improve the geometric reconstruction of 3D anatomy (e.g., pelvic bone) from few 2D X-ray images.

To our knowledge, there is no existing method to reconstruct the 3D mesh models from a single-view 2D medical image, which is a very challenging problem by using the traditional model-driven or statistical techniques. In this paper, we propose a deep learning based data-driven approach to solve this difficult but inspiring problem.

\section{DeepOrganNet}
In this work, we propose an end-to-end deep neural network, named as DeepOrganNet, to generate 3D / 4D surface meshes of multiple organs from single-view medical image projections by learning the optimal deformations upon the best-selected mesh templates. This strategy not only prevents the poor quality of the reconstructed result with coarse voxelization or non-manifold surface mesh (widely existing in previous methods as discussed in Sec.~\ref{sec:dl_methods}), but also preserves fine and smooth surface details for 3D shape generation and visualization. In this section, we introduce main technique components of the DeepOrganNet model including dataset generation, free-form deformation (FFD) on mesh, and organ reconstruction network with the loss functions.

\subsection{Dataset Generation}
\label{sec:data_generation}
3D object reconstruction is one of the most complicated tasks in computer vision / graphics and visualization fields, compared with object classification, segmentation, retrieval, etc., not to mention 3D object reconstruction from a single-view image. Such work usually needs a large-size dataset in support of the deep learning networks to learn the correct mapping from the 2D projection image to the corresponding 3D object. The current 3D shape reconstruction tasks mainly rely on some public large-size and well-established synthetic 3D shape dataset, such as ShapeNets \cite{chang2015shapenet}, ModelNet10, and ModelNet40 \cite{wu20153d}. Additionally, the input image in most tasks is restricted to the 2D projection of a certain object under an identical lighting condition with a uniform background in order to filter out as much unrelated information as possible.

\textbf{Challenges:} Unlike such natural-image-to-3D-object tasks, the proposed organ reconstruction from a single X-ray image (e.g., 3D / 4D-CBCT projection) is more challenging due to following aspects. First, the dataset of 3D organ shapes, such as human lungs (in this paper, we use lung organs as illustrative examples), is very limited, and there exists neither established synthetic dataset nor available clinical dataset with a reasonable scale, which can be adopted in our task. Second, an X-ray image is essentially different from a 2D natural image, which contains obvious object profile and appearance (with the clear background in most previous works); instead the X-ray image contains the structures and details inside the object or occluded from the viewpoint (with the complicated background and noises). Third, the proposed framework does reconstruct multiple objects simultaneously, such as left and right lungs as an example. However, the previous existing approaches could work on reconstructing only one object.

\textbf{Generated Organ Meshes:} Firstly, we address the limited dataset challenge on both 3D lung models and the corresponding 2D medical image projections. We propose a feasible strategy to generate a large number of synthetic 3D-CBCT projection images along with their corresponding 3D two-lung (left and right lungs) surface meshes with various geometric shapes by using a small amount of 3D / 4D phantom data. Given a 3D digital phantom (i.e., a volumetric image) ${I}$, we first apply the Snakes segmentation method~\cite{kass1988snakes} to segment the 3D lung mask images and then extract the isosurface mesh $S$ from the segmented lung mask image by Marching Cubes algorithm~\cite{lorensen1987marching}. After that, we employ a variety of shape deformations and spatial translations on $S$ to get a new mesh $S'$. For the organ shape deformation, we first manipulate the coordinates by multiplying a scaling ratio to globally stretch or compress the lung shape in $S$. The scaling ratios are either constant or gradually change. We then semi-automatically apply local distortions (e.g., dents / concaves, convexes, abnormal parts) on each lung shape using Blender software tool. Both the above global and local deformations are manipulated under the guidance of our collaborative doctors to resemble and cover the real lung shape variations. All the procedures are performed based on 3D / 4D phantoms within different respiration phases to capture the real lung breathing motions. For the organ spatial arrangement, we randomly disturb the distance between each left / right lung's bounding box center and origin within a reasonable range based on the original lung positions in the phantom to resemble various real lung shape cases, and the final new mesh is $S'$.

\textbf{Generated Volumetric Images:} Once we have the deformed surface mesh $S'$ along with the original 3D digital phantom image ${I}$, the deformed 3D digital phantom image ${I}'$ can be computed. Then, we can generate the corresponding single-view (e.g., front-view) 3D-CBCT projections. It is noted that we can obtain the deformation vector field (DVF) $\Delta \mathbf{D}_{V}$ of the mesh vertices from surface $S$ to $S'$ as follows:\vspace{-1mm}
\begin{equation}
\Delta \mathbf{D}_{V}=\mathbf{V}_{S'}-\mathbf{V}_{S},\vspace{-1mm}
\end{equation}
where $\mathbf{V}_{S'}, \mathbf{V}_{S}\in \mathbb{R}^{N \times 3}$ are the positions of mesh vertices in $S'$ and $S$ and $N$ is the number of mesh vertices. As a result, for each voxel $\boldsymbol{\alpha} _{i}$ in $I$, we can estimate its deformation vector $\Delta \mathbf{d}_{\boldsymbol{\alpha}_{i}}$ by incorporating the DVF of its k-nearest neighboring vertices on mesh $S$. Such process can be written as follows:\vspace{-1mm}
\begin{equation}
\Delta \mathbf{d}_{\boldsymbol{\alpha}_{i}}=\mathbf{H}_{\boldsymbol{\alpha}_{i}}\mathbf{J}_{\boldsymbol{\alpha}_{i}}\Delta \mathbf{D}_{V},\vspace{-1mm}
\end{equation}
where $\mathbf{J}_{\boldsymbol{\alpha}_{i}} \in \mathbb{R}^{K \times N}$ is an one-hot encoding matrix for the indices of the $K$ neighbors (in this work, $K$ is 4). $\mathbf{H}_{\boldsymbol{\alpha}_{i}} \in \mathbb{R}^{K}$ is a weight vector in which:\vspace{-1mm}
\begin{equation}
\begin{cases}
h_{\boldsymbol{\alpha}_{i}(\kappa)} =\frac{1}{K}& \text{ if } ||\mathbf{v}_{\kappa}-\boldsymbol{\alpha}_{i}|| \leq \psi,\\
 h_{\boldsymbol{\alpha}_{i}(\kappa)} =\frac{1}{K||\mathbf{v}_{\kappa}-\boldsymbol{\alpha}_{i}||}& \text{ otherwise },\vspace{-1mm}
\end{cases}
\end{equation}
where $\kappa \in 1, ..., K$ and $\psi$ is the norm of the maximum DVF of mesh vertex in $\Delta \mathbf{D}_{V}$. We then can calculate the resulting $I'$ with all voxels' DVF using the reconstruction method with the deformation field map~\cite{ren2008novel,zhong20163d}.

\textbf{Generated 2D Projection Images:} Finally, we apply the Siddon ray tracing algorithm~\cite{siddon1985fast} to generate the desired 2D front-view 3D / 4D-CBCT images of $S'$ by tracing the path of light through voxels in the 3D volumetric image $I'$. To better simulate the realistic raw target CBCT projections from the digital phantom data and test the sensitivity of our method to the realistic complications, after the noise-free ray line integrals are computed according to the above ray tracing method, the noisy signal at each pixel on the CBCT projections is generated based on the noise model with Poisson and normal distributions~\cite{la2005reduction,wang2006penalized,zhong20163d}.

\begin{figure}[h!]
	\begin{center}
		\includegraphics[width=1.0\linewidth]{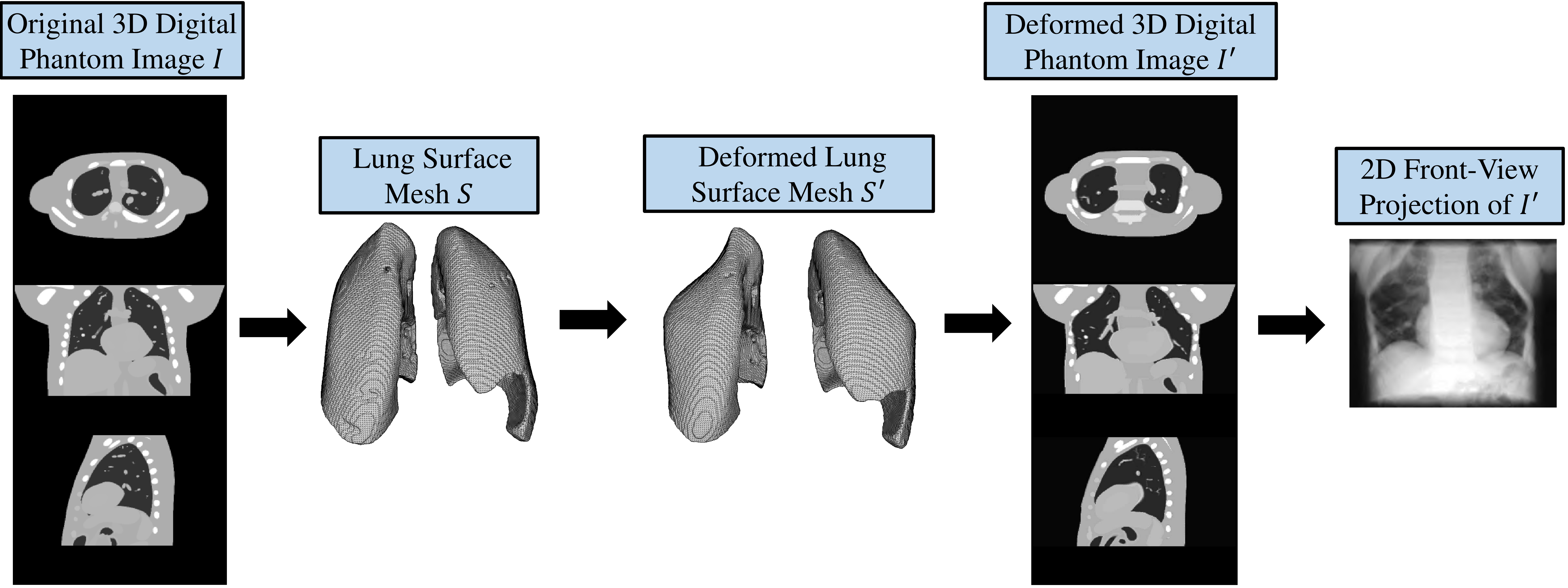}
		\caption{The flowchart of dataset generation.}\vspace{-8mm}
		\label{fig:datagen}
	\end{center}
\end{figure}

In this way, we can generate as many 3D lung meshes and their corresponding 3D / 4D-CBCT medical projections as possible, which is quite crucial for our proposed data-driven deep learning framework. Accordingly, we can cover various types of abnormalities caused by lesions, injuries, or singularities by applying different kinds of global and local deformations in the dataset generation and intentionally increase the ratio of such abnormalities in the dataset to provide our network with adequate prior knowledge to deal with potential unusual cases. Fig.~\ref{fig:datagen} demonstrates the flowchart of the dataset generation. More detailed information is in the experiment section (Sec.~\ref{sec:implement_details}).

\subsection{Free-Form Deformation (FFD) on Mesh}
\begin{figure}[h]
	\begin{center}
		\includegraphics[width=0.7\linewidth]{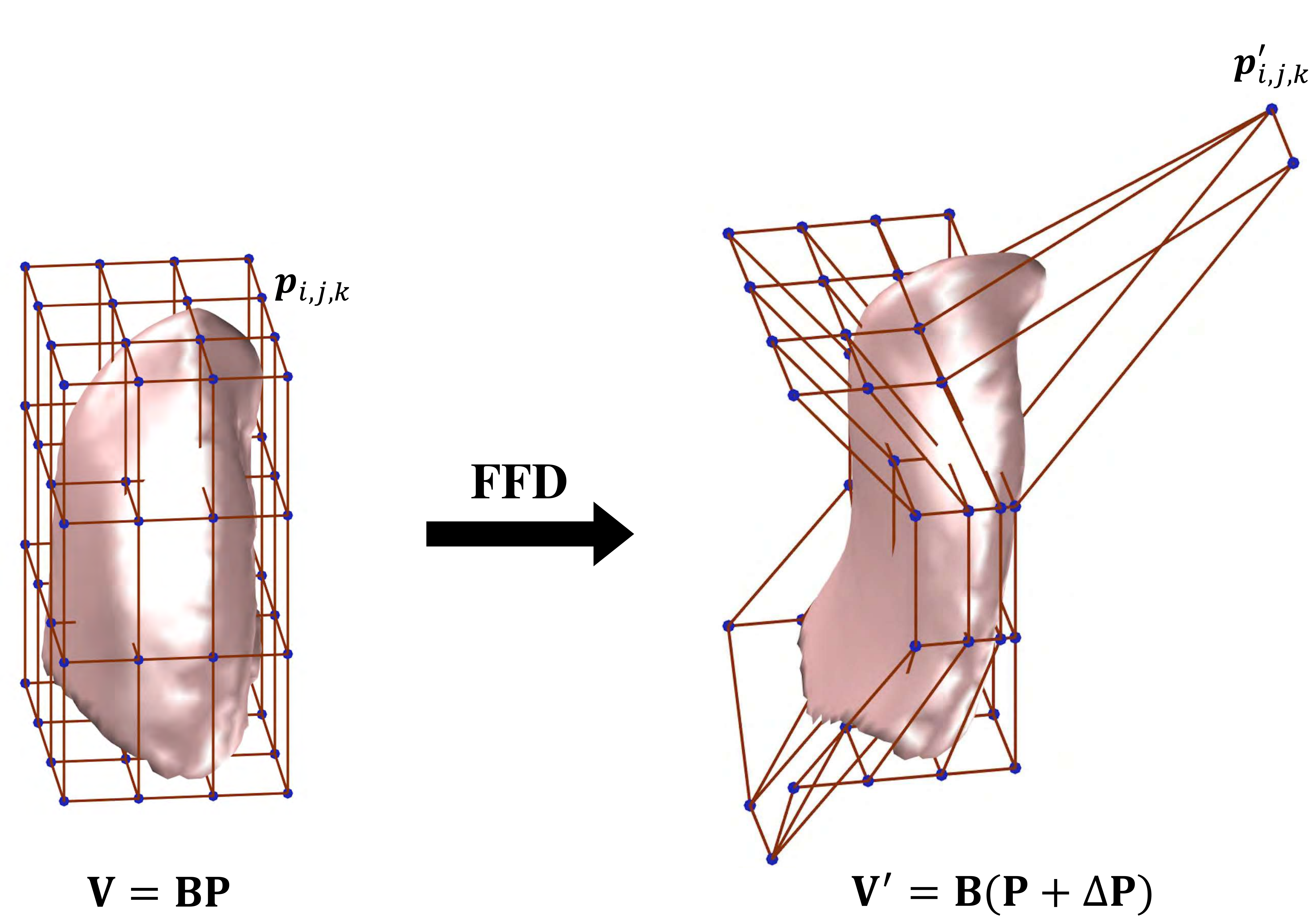}
		\caption{FFD process on a 3D lung shape: it is deformed according to the displacement of the control points on a $4 \times 4 \times 4$ grid.}\vspace{-4mm}
		\label{fig:ffd}
	\end{center}
\end{figure}

A 3D template mesh $\Omega=(\mathbf{V}, \mathbf{F})$ consists of a set of $N$ vertices $\mathbf{V}=\left \{\mathbf{v}_{1},\mathbf{v}_{2},...,\mathbf{v}_{N} \right \}$ and a set of $M$ faces $\mathbf{F}=\left \{\mathbf{f}_{1},\mathbf{f}_{2},...,\mathbf{f}_{M} \right \}$. A high-quality 3D mesh object usually requires dense vertices to represent fine details and thus it is computationally unfriendly, if one intends to deform it pointwisely. Instead, FFD \cite{sederberg1986free} deforms the 3D mesh object through a small amount of control points. FFD introduces a 3D control point grid of size $l \times m \times n$, which encloses the target 3D mesh and performs the deformation in a trivariate tensor-product spline function, where the position of each vertex on the target mesh can be calculated:\vspace{-1mm}
\begin{equation}
\label{eq:ffd}
\mathbf{v}(s,t,u)=\sum_{i=0}^{l}\sum_{j=0}^{m}\sum_{k=0}^{n}B_{i,l}(s)B_{j,m}(t)B_{k,n}(u)\mathbf{p}_{i,j,k},\vspace{-1mm}
\end{equation}
where $\mathbf{v}(s,t,u)$ is an arbitrary mesh vertex coordinate in the coordinate system defined by three orthogonal axes ${s}$, ${t}$, and ${u}$. $B_{p,q}(x)=\binom{p}{q}(1-x)^{p-q}x^{q}$ is a binomial function called Bernstein polynomial of degree $q$, and $\mathbf{p}_{i,j,k}$ is the control point at the node $(i,j,k)$ on the grid. From Eq.~(\ref{eq:ffd}), we notice that the vertex placement of the target mesh is essentially a weighted sum of the control points. Denote $\mathbf{V}\in \mathbb{R}^{N\times 3}$ as the matrix form of vertices on mesh $\Omega$, then the mesh vertex representation can be converted:\vspace{-1mm}
\begin{equation}
\mathbf{V}=\mathbf{B}\mathbf{P},\vspace{-1mm}
\end{equation}
where $\mathbf{B} \in \mathbb{R}^{N\times L}$ is the matrix form of the trivariate Bernstein tensor for all $N$ vertices and $\mathbf{P}\in \mathbb{R}^{L\times 3}$ is the matrix form of control point coordinates. Suppose given the displacement of these control points $\bigtriangleup\mathbf{P}$, the corresponding deformed mesh $\Omega'=(\mathbf{V}', \mathbf{F})$ in which $\mathbf{V}'$ is:\vspace{-1mm}
\begin{equation}\label{ha}
\mathbf{V}'=\mathbf{B}(\mathbf{P}+\bigtriangleup\mathbf{P}).\vspace{-1mm}
\end{equation}
As shown in Fig.~\ref{fig:ffd}, in this way, the objective of the proposed deep learning network (DeepOrganNet) is to infer a $\mathbf{\bigtriangleup P}$ such that the resulting mesh $S'$ best matches the shape of 3D lung organ surface according to the input 2D X-ray image.\vspace{-4mm}

\begin{figure*}[htb]
	\begin{center}
		\includegraphics[width=0.95\linewidth]{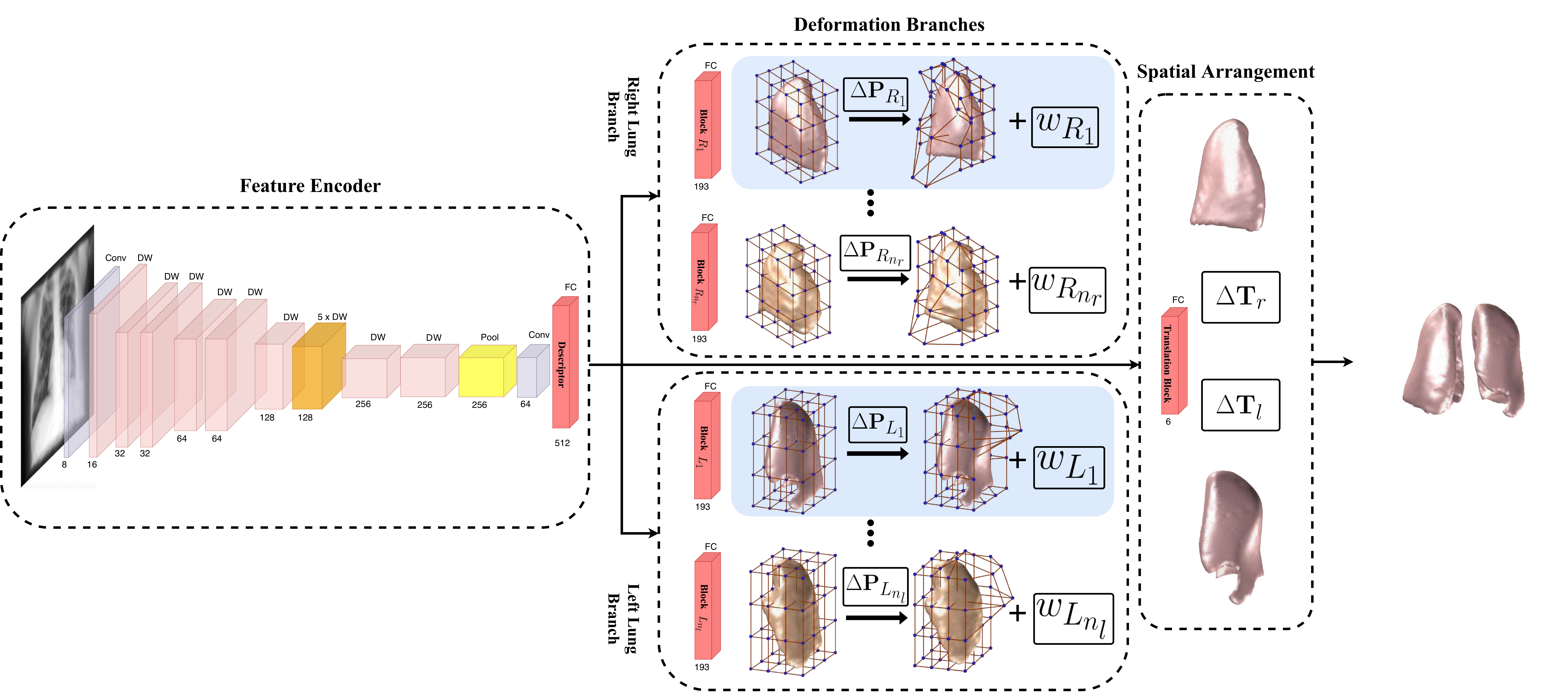}
		\caption{The architecture of our DeepOrganNet. The DeepOrganNet first encodes the input image into a descriptor using MobileNets (without fully-connected layers) followed by a $1 \times 1$ convolution layer (dimension reduction). DW refers to the depthwise separable convolution block (two separable convolutional layers, functionally equivalent to a regular convolutional layer) and the numbers are output channel sizes (i.e., widths) of each layer / block. Every template in either left or right lung branch learns its own selection weight $w$ and deformation parameters $\bigtriangleup\mathbf{P}$ through an independent fully-connected layer with dimension 193, including 192 for $\bigtriangleup\mathbf{P}$ and 1 for $w$. The deformed templates with the highest selection weights (e.g., templates $L_{1}$ and $R_{1}$ are selected in this example) in both branches are arranged according to the translation vector $\bigtriangleup\mathbf{T}_{l}$ and $\bigtriangleup\mathbf{T}_{r}$ learned from another fully-connected layer to generate the final combined multi-organ meshes.}\vspace{-8mm}
		\label{fig:archi}
	\end{center}
\end{figure*}

\subsection{Organ Reconstruction Network}
\subsubsection{Network Architecture Design}
The pipeline of DeepOrganNet framework consists of three functional components: feature encoder block, deformation block, and spatial arrangement block. The overall architecture is shown in Fig.~\ref{fig:archi}.

Given a single-view 3D / 4D-CBCT projection image, our network first encodes it into a latent descriptor, which contains effective information for different purposes in the reconstruction stage. Due to the dataset availability and the specific training objective, the image encoder should be lightweight to alleviate the overfitting risk but still quite efficient for the reconstruction task. Jack et al. \cite{jack2018learning} has justified the adequate ability for MobileNet in the intra-class deformation, so we apply and fine-tune the pre-trained MobileNets \cite{howard2017mobilenets} to encode the input medical image. MobileNets are compact and Inception style \cite{szegedy2016rethinking} network, which factorizes a standard convolution into a depthwise convolution and a $1 \times 1$ pointwise convolution. In addition, MobileNets introduce a width multiplier, which can reduce the width for each layer by a constant ratio and thus give us more freedom to adjust our network to best fit a relatively small amount of data in medical image scenarios. In our work, we only adopt the convolution layers in MobileNets and add a $1 \times 1$ convolutional layer after that to generate the image descriptor with a reasonable dimension. The detailed network configuration is shown in Fig.~\ref{fig:archi}. Through our extensive experiments in Sec.~\ref{sec:experiments}, we find that the lightweight MobileNets is sufficient and robust to extract the informative features from a single-view medical image with complicated background and noises. Our input image is quite different from the one in most of the current natural-image-to-3D-object tasks, since their inputs are 2D images with clear object profile and boundary, which are generated based on the light illumination and reflection; however, our X-ray images are generated based on ray tracing technique to compute the attenuation of the energy absorption. One of the advantages in the X-ray-image-to-3D-object task is that the input X-ray images can include some shape information, which is occluded by the natural 2D images. It can make the viewer to see through the front shape surface and thus alleviate the occlusions. We will show examples in the experiment section (Sec.~\ref{sec:experiments}).

Furthermore, another advantage of our DeepOrganNet, compared with current natural-image-to-3D-object tasks in which the prediction is only designed for a single object, our task is essentially designed for reconstruction of multiple separate objects, along with a spatial arrangement between each other. As a result, we split different branches from the whole image descriptor to reconstruct different organs (i.e., different disconnected components), such as the left and right lungs independently, instead of learning all the objects together. Based on our observations and explorations, this scheme is more effective for the network to learn discriminative features from different objects than the one to learn everything together (such as using one branch scheme). Each branch is responsible for the corresponding single lung generation by deforming differently-geometric templates. In this work, we use left and right lung organs as the testbed for our study, but the proposed framework can be easily extended to multi-organ (more than two organs) scenarios. Suppose we select $n_{l}$ left lung templates and $n_{r}$ right lung templates for corresponding branches, left lung branch learns a set of the deformation parameters for all of the $n_{l}$ templates $\left \{ \bigtriangleup\mathbf{P}_{L_{i}} \right \}_{i=1}^{n_{l}}$ according to the left lung shapes from the input 2D image simultaneously, where $\bigtriangleup\mathbf{P}_{L_{i}}\in \mathbb{R}^{l \times m \times n \times 3}$ is the deformation parameter (i.e., the control points' displacements) for a single template $L_{i}=(\mathbf{V}_{L_{i}},\mathbf{F}_{L_{i}})$ and $l \times m \times n$ is the number of control points. The deformed template ${L}'_{i}$ can be achieved by:\vspace{-1mm}
\begin{equation}
\label{eq:deformed_v}
{\mathbf{V}'}_{{L}'_{i}}=\mathbf{B}_{L_{i}}\left(\mathbf{P}_{L_{i}} +\bigtriangleup\mathbf{P}_{L_{i}} \right),
\end{equation}
\begin{equation}
\label{eq:deformed_f}
\mathbf{F}'_{{L}'_{i}}=\mathbf{F}_{L_{i}},\vspace{-1mm}
\end{equation}
where $\mathbf{B}_{L_{i}}$ and $\mathbf{P}_{L_{i}}$ are the pre-computed transformation matrix and control point position matrix for the template $L_{i}$. The mesh connectivity does not change during the deformation as shown in Eq.~(\ref{eq:deformed_f}). In addition, the left lung branch also learns a set of selection weights $\left \{w_{L_{i}} \right \}_{i=1}^{n_{l}}$ for each left lung template, which are defined along with the template deformations as shown in Eq.~(\ref{eq:deform_loss}) and Eq.~(\ref{eq:regular_loss}). The final left lung prediction is then selected by:\vspace{-1mm}
\begin{equation}
L_{pred}={L}'_{i_{max}},\vspace{-1mm}
\end{equation}
where $i_{max}=\arg \max\left \{ w_{L_{i}} \right \}_{i=1}^{n_{l}}$. Similarly, right lung branch applies the same procedure as the left lung branch to obtain the final right lung prediction $R_{pred}$. By splitting two branches to extract the corresponding effective information from the image descriptor, the learning objectives become more specific and clearer. At this stage, the network only focuses on how to deform the templates with respect to the lung geometry from the input image. The spatial information, such as the gap / distance and the relative positions between left and right lungs, are reserved for the next stage.

As long as we have $L_{pred}$ and $R_{pred}$ ready, the next step is to combine them together so as to generate final left and right lung meshes with the correct relative spatial arrangement according to the input image. In order to achieve this, we learn two translation vectors $\bigtriangleup \mathbf{T} _{l}$, $\bigtriangleup \mathbf{T} _{r}$ from the image descriptor. Then the entire prediction of the new organ meshes $\Omega'=(\mathbf{V}'_{\Omega'},\mathbf{F}'_{\Omega'})$ of both lungs is:\vspace{-1mm}
\begin{equation}
\mathbf{V'}_{\Omega'}=\left \{\mathbf{V'}_{L_{pred}}+\bigtriangleup \mathbf{T} _{l},\mathbf{V'}_{R_{pred}}+\bigtriangleup \mathbf{T} _{r}\right \},
\end{equation}
\begin{equation}
\mathbf{F'}_{\Omega'}=\left \{\mathbf{F'}_{L_{pred}},\mathbf{F'}_{R_{pred}}\right \}.
\end{equation}

\subsubsection{Loss Functions}
In this subsection, we define three kinds of losses in our network not only to constrain the output shape results but also to optimize the training process.

\textbf{Deformation Loss:} To ensure the deformation accuracy, we choose Chamfer loss \cite{fan2017point} to regulate the accuracy of the vertex locations on a single lung prediction. In general, the Chamfer loss is defined as:\vspace{-1mm}
\begin{equation}
\label{eq:cd}
C(\mathbf{P},\mathbf{Q})=\sum_{\mathbf{p \in \mathbf{P}}} \min_{\mathbf{q} \in \mathbf{Q}}\left \| \mathbf{p}-\mathbf{q} \right \|_2^2+\sum_{\mathbf{q} \in \mathbf{Q}} \min_{\mathbf{p} \in \mathbf{P}}\left \| \mathbf{p}-\mathbf{q} \right \|_2^2,\vspace{-1mm}
\end{equation}
where $\mathbf{p}$ and $\mathbf{q}$ are vertices from two different point sets from meshes $\mathbf{P}$ and $\mathbf{Q}$. Essentially, for each point in $\mathbf{P}$ or $\mathbf{Q}$, the Chamfer loss finds the nearest vertex in the other point set and sums up all the pair-wise distances. In our framework, we apply weighted Chamfer loss for both lung branches as:\vspace{-1mm}
\begin{equation}
\label{eq:deform_loss}
 \begin{aligned}
\mathfrak{L}_{deform}=\sum_{i=1}^{n_l}w_{L_{i}}C(\mathbf{V'}_{L_{pred}}, \mathbf{V}_{L_{gt}}) + \\
\sum_{i=1}^{n_r}w_{R_{i}}C(\mathbf{V'}_{R_{pred}},\mathbf{V}_{R_{gt}}),\vspace{-1mm}
\end{aligned}
\end{equation}
where $\mathbf{V}_{L_{gt}}$ and $\mathbf{V}_{R_{gt}}$ are the ground truth for left and right lung meshes (aligned at the origin), respectively. In this way, the proposed network is enforced to give the highest weight to the template, which can be deformed best to match the ground truth. Now, we can select the best template among all potential candidates in the datasets for predicting each organ individually and automatically.

\textbf{Translation Loss:} The second loss term $\mathfrak{L}_{trans}$ is intended to learn the translation vectors $\bigtriangleup \mathbf{T} _{l}$ and $\bigtriangleup \mathbf{T} _{r}$. It is defined as:\vspace{-1mm}
\begin{equation}
\mathfrak{T}_{trans}=\left \| \bigtriangleup \mathbf{T}_l-\mathbf{ctr}_{l} \right \|_2^2+\left \| \bigtriangleup \mathbf{T}_r-\mathbf{ctr}_{r} \right \|_2^2,\vspace{-1mm}
\end{equation}
where $\mathbf{ctr}_l$ and $\mathbf{ctr}_r$ are the ground truth translation vectors (i.e., two global translation vectors between the origin and the bounding box centers of left and right lungs in all ground truth meshes).

\textbf{Regularization Loss:} Our network deforms all templates according to input 2D images. Sometimes, the reconstruction results are achieved by tremendous deformations from a template that is not the closest one in the template pool. We introduce a weight regularization term similar to the one in~\cite{jack2018learning} to encourage the network to give higher weight to the template closer to the ground truth. In this way, the overall performance of the network becomes more rational and intuitive:\vspace{-1mm}
\begin{equation}
\label{eq:regular_loss}
\mathfrak{L}_{w}=\sum_{i=1}^{n_l}w_{L_{i}}\left \| \bigtriangleup\mathbf{P}_{L_{i}} \right \|_2^2+\sum_{i=1}^{n_r}w_{R_{i}}\left \| \bigtriangleup\mathbf{P}_{R_{i}} \right \|_2^2,\vspace{-2mm}
\end{equation}
where this loss is defined on the deformations of the control points.

The \textbf{total loss} is a weighted sum of all the above three kinds of losses as follows:\vspace{-1mm}
\begin{equation}
\mathfrak{L}_{total}=\mathfrak{L}_{deform}+\lambda_1 \mathfrak{L}_{trans} + \lambda_2\mathfrak{L}_{w},\vspace{-2mm}
\end{equation}
where $\lambda_1 =50$ and $\lambda_2 =1$ in experimental settings, which are determined based on the corresponding order of the magnitude and balanced by the optimal network performance via our extensive experiments.

It is worth mentioning that through the strategy of integrating the deformation weights in the loss function, the proposed DeepOrganNet can automatically select the proper templates so that the network has a good prior information to start with for each organ. The risk of non-manifold issue in the reconstructed shape meshes, such as~\cite{fan2017point,wang2018pixel2mesh,Smith2019}, is dramatically alleviated. In addition, FFD deforms the templates with a small amount of control points compared with vertex-wise deformation, e.g., $64$ vs $10$K deformation parameters, which is quite efficient. Furthermore, FFD can realize the high-order interpolation for the deformation computations, so that the mesh surface smoothness is well maintained and no additional loss term as in~\cite{wang2018pixel2mesh,Smith2019} is required beyond the Chamfer loss (fidelity term) to yield a good inference.

\section{Implementation Details}
\label{sec:implement_details}
In this section, we introduce our dataset preparation and network training details followed by evaluation metrics which we use to measure the experimental results.

\textbf{Dataset Preparation:} In order to evaluate the proposed DeepOrganNet, we use following phantoms, patient studies in lung imaging and motion datasets. There are two 3D / 4D digital phantoms, i.e., a dynamic NURBS-based cardiac-torso (4D NCAT) phantom (4D images and motions are provided) and 4D extended cardiac-torso (XCAT) \cite{dissertation:Segars2001}, being used as basis models to generate a reasonable number of 3D lung surface meshes and corresponding 3D / 4D-CBCT projections. They both have 10 breathing phases in 3D volumetric images (e.g., $256 \times 256 \times 150$ with voxel size of $1 mm \times 1 mm \times 1 mm$). We generate 542 pairs of (left and right) lungs with various shapes and different spatial arrangements together with their corresponding 2D single front-view CBCT projections (see Sec.~\ref{sec:data_generation}). We use the first five phases of NCAT and XCAT of 4D-CBCTs to build our training dataset and leave the rest for testing evaluation purpose. All two-lung models are normalized along the sagittal axis and translated to the origin. The bounding box size of all these models is within $1.35 \times 1.25 \times 1$ along the transverse, coronal, and sagittal axes. We then compute the bounding box centers of left and right lungs in the two-lung meshes and translate them to the origin to form the ground truth for the two deformation branches. The input 2D front-view CBCT projections are grayscale images of size $192 \times 256$ with pixel size of $1 mm \times 1 mm$. In the experiments, we randomly split the dataset by 446 pairs for training and 96 pairs for testing, respectively. We also test our model performance on deformable image registration (DIR) Lab (ten lung cancer patient 4D-CT datasets with ten respiration phases each)~\cite{Castillo:2009}, Japanese Society of Radiological Technology (JSRT) database (247 chest X-ray images)~\cite{shiraishi2000development} for lung shape reconstruction.

\textbf{Training Details:} Our task is to generate left and right lung shapes from an image with noisy background and limited dataset, in order to reach a good balance between the prediction accuracy and the network overfitting risk. We set the MobileNets~\cite{howard2017mobilenets} (pre-trained on ImageNet dataset~\cite{deng2009imagenet}) width multiplier to be 0.25 and the width (i.e., channel number) for each layer is shown in Fig.~\ref{fig:archi}. For each lung branch in the network, we have two single lung templates from XCAT and NCAT, respectively. The 3D control point grids for every template from two branches are set to be $4 \times 4 \times 4$, which yields to 64 control points per template. We train the network for 65K steps using Adam optimizer with learning rate as $1\times 10^{-3}$. The batch size is 32. The total training time is 4 hours on a single Nvidia GTX 1080 GPU with 8 GB GDDR5X.

\textbf{Evaluation Metrics:} Following the standard 3D shape reconstruction evaluation method, we use five different kinds of numeric metrics to evaluate the performance of our model and compare with the existing state-of-the-art techniques.

\emph{Chamfer distance (CD)} is applied in both training and testing processes. The formal expression is shown in Eq.~(\ref{eq:cd}). It measures bidirectional overall vertex-wise distance between two meshes.

\emph{Earth mover's distance (EMD)} \cite{rubner2000earth} is designed to compute the minimal sum of distance over all possible one-to-one mapping between points in $\mathbf{P}$ and points in $\mathbf{Q}$, where $\mathbf{P}$ and $\mathbf{Q}$ are two point sets of the same size. The EMD can be written as:\vspace{-1mm}
\begin{equation}
EMD(\mathbf{P},\mathbf{Q})=\underset{\phi :\mathbf{P}\rightarrow \mathbf{Q}}{\min}\sum_{\mathbf{p} \in \mathbf{P}}\left \| \mathbf{p}-\phi (\mathbf{p}) \right \|_2,\vspace{-2mm}
\end{equation}
where $\phi$ is a bijection from $\mathbf{P}$ to $\mathbf{Q}$.

\emph{Hausdorff distance (HD)} is adopted to measure the largest inconsistency between the reconstruction result and the ground truth. A lot of previous 3D reconstruction works did not list it as their evaluation metric because they mainly focused on point cloud reconstruction, which is insensitive to small amount of outliers. In geometric modeling and computer graphics, Hausdorff distance is a widely-used indicator to check the reconstructed mesh quality since even small amount of outliers may undermine the mesh surface consistency and quality, especially for visualization and rendering. In our experiments, we measure the Hausdorff distance between prediction and ground truth with respect to both points and surface meshes~\cite{cignoni1998metro}. Suppose $\mathbf{p}$ and $\mathbf{q}$ are the points sampled from point clouds (or surface meshes) of $\mathbf{P}$ and $\mathbf{Q}$ accordingly, the HD in terms of point clouds (or surface meshes) can be written as:\vspace{-1mm}
 \begin{equation}
 \footnotesize
 	HD(\mathbf{P},\mathbf{Q})=\textup{max}\left [ \underset{\mathbf{p}\in \mathbf{P}}{\textup{max}}\left ( \underset{\mathbf{q}\in \mathbf{Q}}{\textup{min}} \left \| \mathbf{p}-\mathbf{q} \right \|_2\right ),\underset{\mathbf{q}\in \mathbf{Q} }
 {\textup{max}}\left (\underset{\mathbf{p}\in \mathbf{P}}{\textup{min}} \left \| \mathbf{q}-\mathbf{p} \right \|_2\right )\right ].\vspace{-1mm}
 \end{equation}

\emph{F-score} \cite{wang2018pixel2mesh} is used as the harmonic mean of precision and recall regarding how many points in prediction or ground truth can find the nearest neighbor from the other within a threshold ($\epsilon$). We set $\epsilon=0.001$ in our experiments.

\emph{Intersection over union (IoU)} is used to examine the volumetric similarity between the voxelized prediction and ground truth. The IoU is defined as:\vspace{-1mm}
\begin{equation}
IoU(\mathbf{P},\mathbf{Q})=\frac{|\mathbf{P}\cap \mathbf{Q}|}{|\mathbf{P}\cup \mathbf{Q}|},\vspace{-1mm}
\end{equation}
where $\mathbf{P}$ and $\mathbf{Q}$ are the voxelized 3D models.

Among the above five metrics, for CD, EMD and HD, the smaller the better; while for F-score and IoU, the larger the better.

\section{Experiments}
\label{sec:experiments}
In this section, we conduct extensive experiments of our model on the 3D and 4D synthetic data as well as real patient data. The results are qualitatively and quantitatively compared with several state-of-the-art in deep learning based 3D shape generation (from a single-view image) methods and the traditional reconstruction method. It is noted that for comparison experiments, best results in tables are shown in bold font.

\subsection{3D Lung Shape Reconstruction from Synthetic Images}
\begin{figure}[t!]
	\begin{center}
		\includegraphics[width=1.0\linewidth]{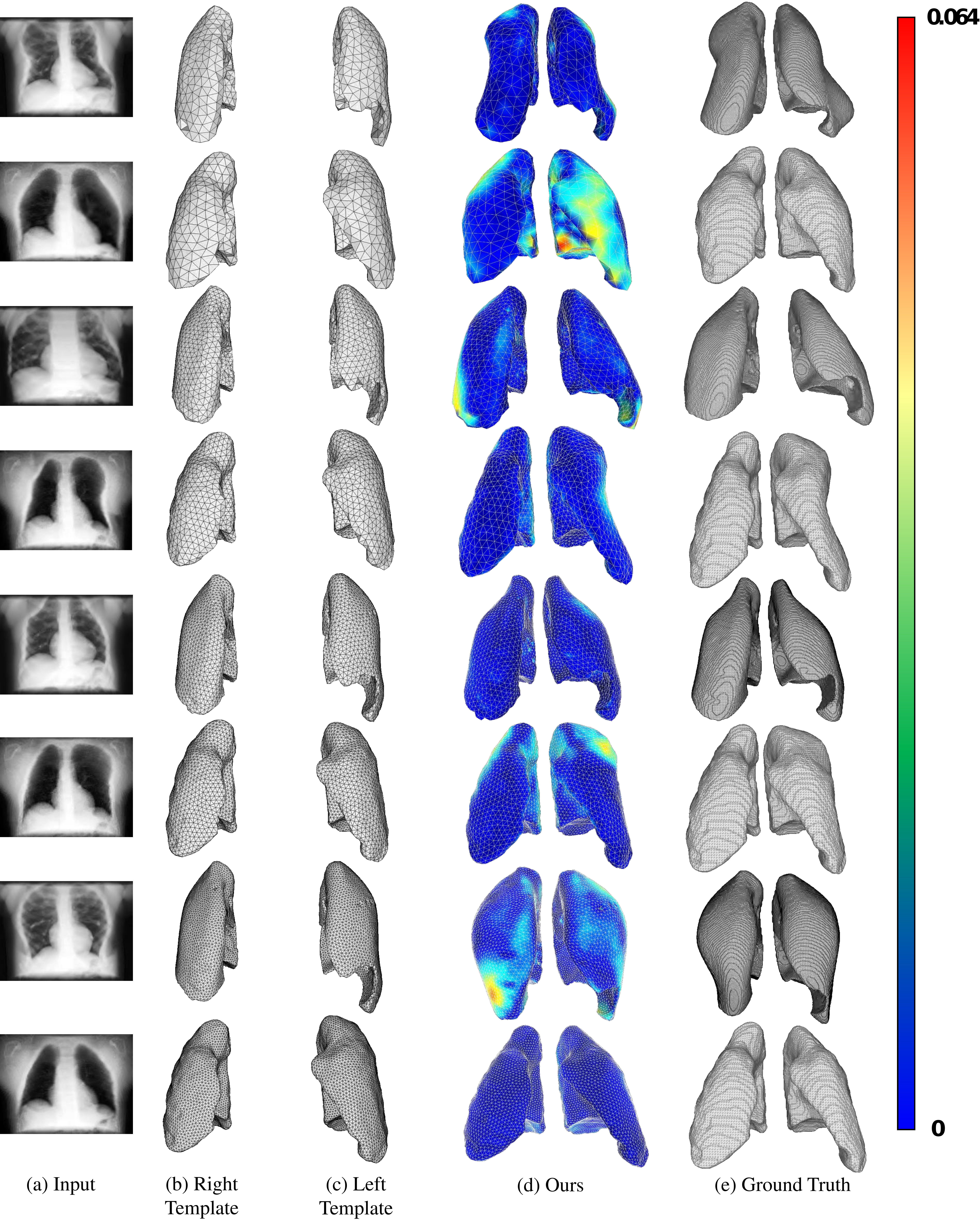}\vspace{-2mm}
		\caption{Qualitative reconstruction and visualization results of some lung shapes with drastic variations. The reconstruction error (HD on mesh) is mapped into a unified colormap range (hotter colors indicate larger errors and colder colors indicate smaller errors) and the mesh resolution increases from top to bottom (e.g., 1K, 2.5K, 5K, 10K vertices).}\vspace{-4mm}
		\label{fig:self}
	\end{center}
\end{figure}

Fig.~\ref{fig:self} shows the reconstruction results of 3D lungs with different shapes based on the synthetic data. Our network is capable of dealing with drastic variations (even though the real-world medical scenarios are far less challenging). For each input image, our network is able to pick the template which most resembles the ground truth model within the corresponding branches, and predict the accurate spatial arrangement between left and right lungs to generate the final high-fidelity 3D lung shape pairs. The reconstruction error (HD on mesh) is mapped into a unified colormap range and it shows that the reconstruction results are pretty good qualitatively and quantitatively.

Since our network learns the deformation parameters which are essentially applied on the control points instead of directly on the template model surface, one can generate the final 3D mesh models with arbitrary resolutions in real-time (e.g., 1K vertices: 20 ms, 2.5K vertices: 21 ms, 5K vertices: 22 ms, 10K vertices: 25 ms) according to the users' needs without re-training the network.

\subsection{Comparison with Deep Learning Based Methods}
\textbf{Comparison with Pixel2Mesh~\cite{wang2018pixel2mesh}:} Intuitively, according to the Pixel2Mesh (P2M) experiment setting, we first use their network to predict a single lung from a single-view input image.

We train P2M on our synthetic dataset following their training details. The quantitative evaluation result is shown in Tab.~\ref{table:p2m}. We fairly set our output mesh vertex number to be the same as the output of P2M network (i.e., 2466). The CD and EMD are both computed between the uniformly sampled 1024 points from the prediction and ground truth such that the comparison can be made not only between the single lung reconstruction from P2M and our network, but also between the lung pair reconstruction (see Tab.~\ref{table:psgn-tab}) and single lung reconstruction of our network. From Tab.~\ref{table:p2m}, our network outperforms P2M on all metrics.
\begin{table}[t!]
	\vspace{-2mm}\caption{Quantitative comparison between P2M and our method on our synthetic dataset.} \vspace{-2mm}
\begin{adjustbox}{width=\columnwidth,center}
	\begin{tabular}{|c|c|c|c|c|c|}
		\hline
		Method & CD     & EMD     & F-score ($\epsilon$ / $1.5\epsilon$)         & IoU    & HD (Mesh) \\ \hline\hline		
		P2M (Left Lung)   & 2.4609 & 76.2620 & 0.5983 / 0.7799 & 0.7190 & 0.1300    \\ \hline
        Ours (Left Lung)  & \textbf{1.7018} & \textbf{57.0856} & \textbf{0.7293 / 0.8910} & \textbf{0.8352} & \textbf{0.0672}    \\ \hline
		P2M (Right Lung)  & 2.3399 & 69.7205 & 0.6111 / 0.8014 & 0.7661 & 0.1022    \\ \hline
        Ours (Right Lung) & \textbf{1.7300}   & \textbf{59.9497} & \textbf{0.7293 / 0.8892} & \textbf{0.8423} & \textbf{0.0786}    \\ \hline
	\end{tabular}\vspace{-6mm}
\end{adjustbox}
\label{table:p2m}
\end{table}

\begin{figure}[t!]
	\begin{center}
		\includegraphics[width=0.65\linewidth]{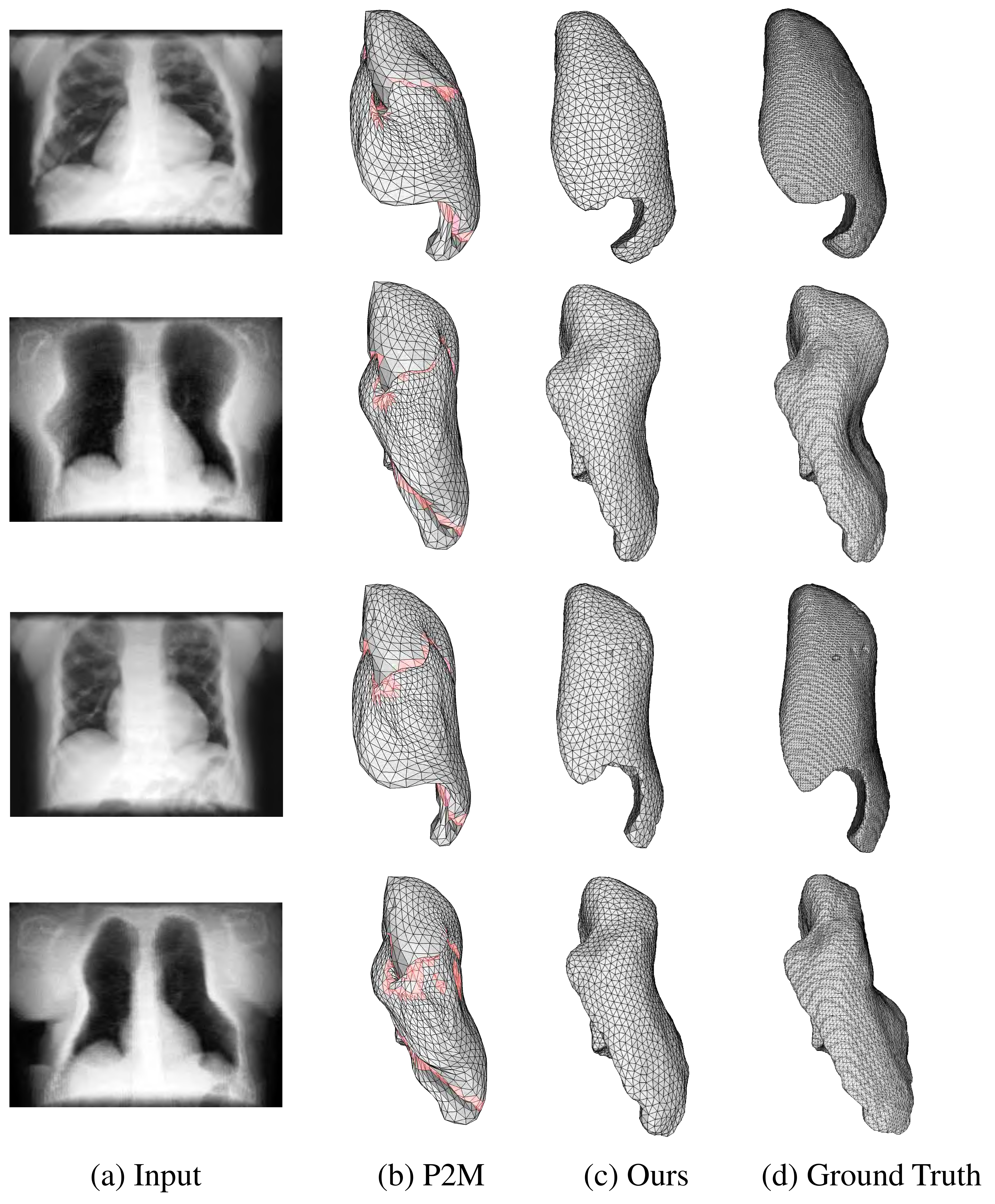}\vspace{-2mm}
		\caption{Qualitative comparison with P2M and our method on left lung model. Our results generate meshes with no non-manifold issue, while the results from P2M have self-intersections (highlighted in red).}\vspace{-7mm}
		\label{fig:left}
	\end{center}
\end{figure}

\begin{figure}[t!]
	\begin{center}
		\includegraphics[width=0.65\linewidth]{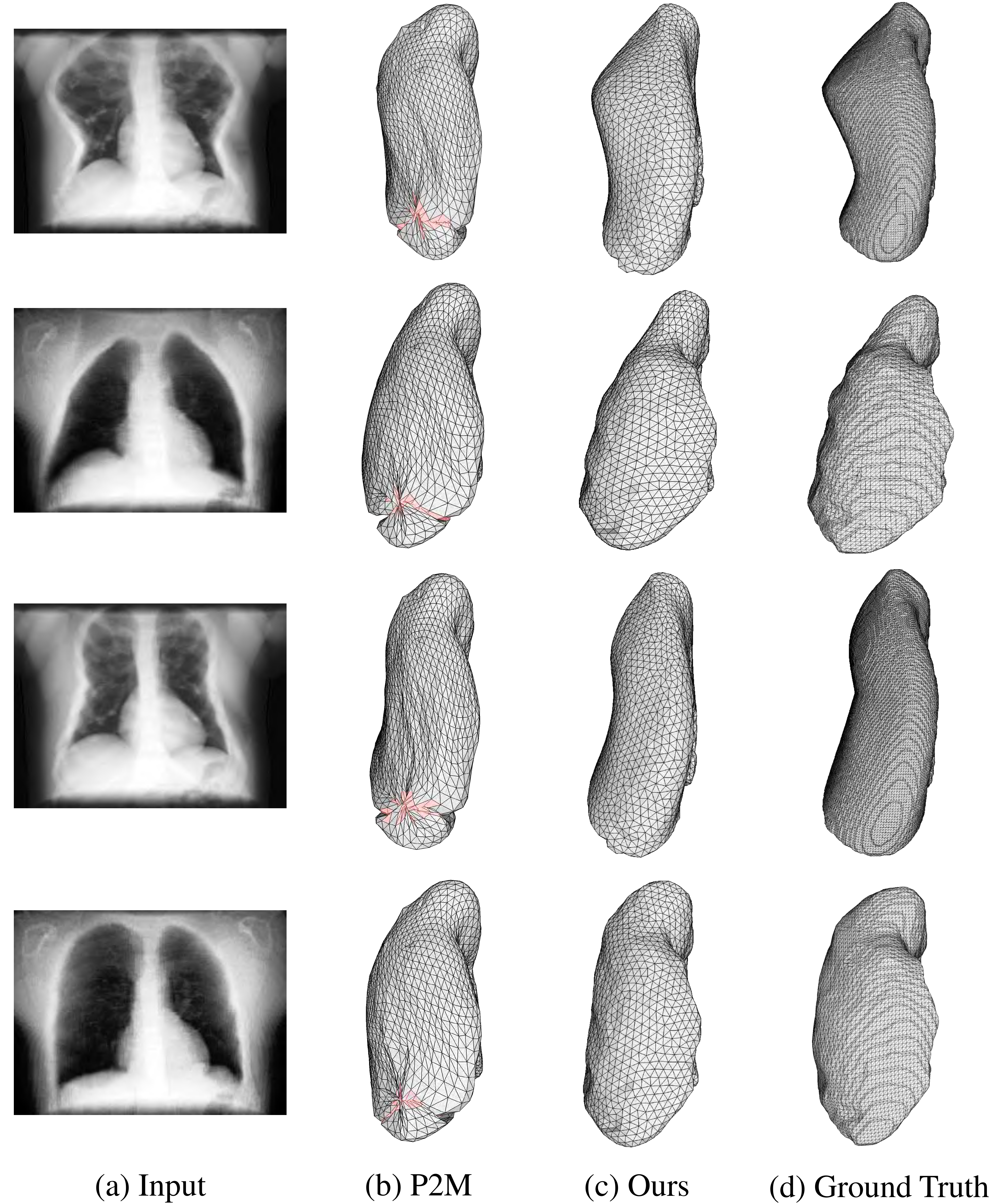}\vspace{-2mm}
		\caption{Qualitative comparison with P2M and our method on right lung model. Our results generate meshes with no non-manifold issue, while the results from P2M have self-intersections (highlighted in red).}\vspace{-7mm}
		\label{fig:right}
	\end{center}
\end{figure}

We also present the qualitative comparison results in Fig.~\ref{fig:left} and Fig.~\ref{fig:right}. Both (P2M and our) networks yield predictions of smooth surface but our model performs better in well preserving the mesh surface geometry without non-manifold issue. The reason may be that the input single-view 3D-CBCT projection image is essentially different from a 2D natural input image. P2M has no mechanism to deal with the ambiguity caused by such ill-posed problem like PSGN. However, our network have more specific templates to start with so as to rule out some uncertainties or local minima, while the initial ellipsoid template in P2M network is too general for this task; and how to modify their network to fit for an initial lung shape is beyond the scope of this work.

We also attempt to infer two lungs together with P2M network by replacing the single ellipsoid with a pair of two ellipsoids. However, the network tends to fuse two separate lungs as a single object. The original weights for each regularization terms need to be further determined to reach a good performance. It seems to be non-trivial to extend P2M framework to multi-object reconstruction scenario.

\textbf{Comparison with Point Set Generation Network~\cite{fan2017point}:} We use our synthetic dataset (i.e., 542 pairs of left and right lungs) with the same training and testing splits to train the Point Set Generation Network (PSGN)~\cite{fan2017point} as our model (discussed in Sec.~\ref{sec:implement_details}) and generate meshes from corresponding prediction point clouds using Ball Pivoting Algorithm~\cite{bernardini1999ball}. Since our model can generate meshes with arbitrary densities, we set the output mesh vertex number as 1024 to fairly compare HD with the mesh generated from PSGN predictions. The CD and EMD are both computed between the prediction and uniformly sampled 1024 points from the ground truth (denser isosurface meshes).

\begin{figure*}[th!]
	\begin{center}
		\includegraphics[width=0.75\linewidth]{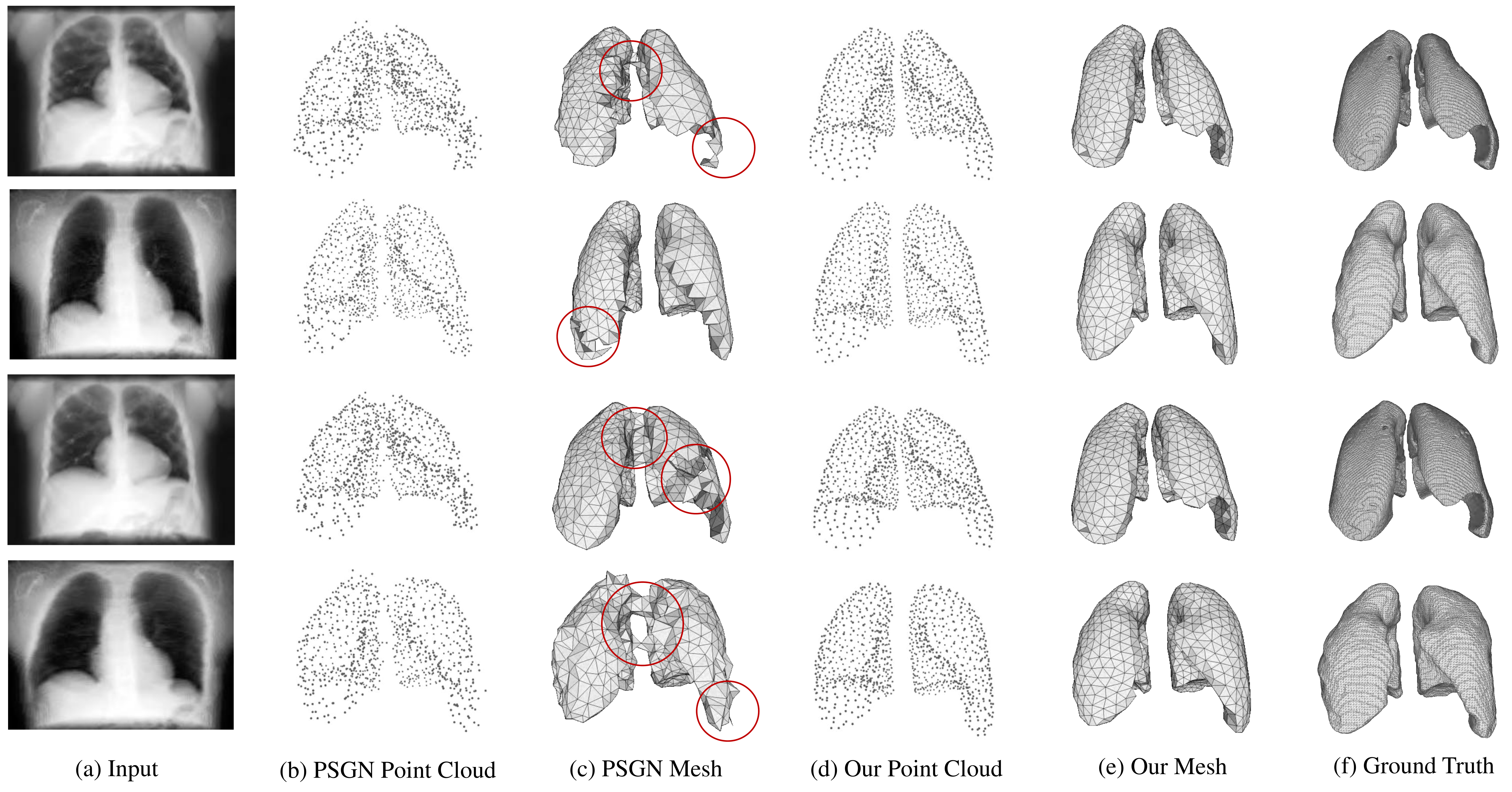}
		\caption{Qualitative comparison between PSGN and ours. Both point clouds and solid surface meshes are given. The failure parts (e.g., self-intersecting triangles, holes, non-manifold triangles, etc.) of PSGN meshes are red-cycled.}\vspace{-8mm}
		\label{fig:psgn}
	\end{center}
\end{figure*}

Tab.~\ref{table:psgn-tab} shows the quantitative evaluation of six different metrics and Fig.~\ref{fig:psgn} provides the qualitative comparison. Our network outperforms PSGN in most metrics. In terms of point-wise HD and F-score($\epsilon$) evaluation, PSGN tends to get slightly better numeric results since the PSGN generates points independently, thus it has more degrees of freedom. However, the EMD of PSGN is much larger since the point cloud inference from PSGN is irregularly distributed, and sometimes the points from one lung are much more than those from the other. Although PSGN has comparable performance in most of the point-wise evaluation metrics, it does not guarantee a high-quality 3D surface mesh. The mesh-based HD is nearly $50\%$ higher than ours since there are a lot of meshing failures (e.g., self-intersecting triangles, holes, non-manifold triangles, etc.) and bumpy details in the generated surface meshes. In addition, PSGN learns a lung pair as a single object, when the gap between two lungs is small, the (post-processing) meshing algorithm is difficult to separate them.
\begin{table}[htb!]
	\caption{Quantitative comparison between PSGN and our method on our synthetic dataset.}  \vspace{-2mm}
    \setlength{\tabcolsep}{1.0mm}
    \centering
    \label{table:psgn-tab}
	\begin{adjustbox}{width=\columnwidth,center}
		
		\begin{tabular}{|c|c|c|c|c|c|c|}
			\hline
			\rule{0mm}{2.2mm} Method & CD            & EMD            & F-score ($\epsilon$ / $1.5\epsilon$)         & HD (Point)         & IoU             & HD (Mesh)       \\ \hline\hline	
			\rule{0mm}{2.2mm} PSGN & 3.0122        & 186.8821       & \textbf{0.4384} / 0.6377 & \textbf{0.0960} & 0.8002          & 0.1491          \\ \hline
            \rule{0mm}{2.2mm} Ours & \textbf{2.8955} & \textbf{70.7083} & 0.4375 / \textbf{0.6650} & 0.0980          & \textbf{0.8148} & \textbf{0.1000} \\ \hline
		\end{tabular}
	\end{adjustbox}\vspace{-4mm}
\end{table}

\subsection{Comparison with Traditional Reconstruction Method}
\begin{figure}[h!]
	\begin{center}
		\includegraphics[width=0.95\columnwidth]{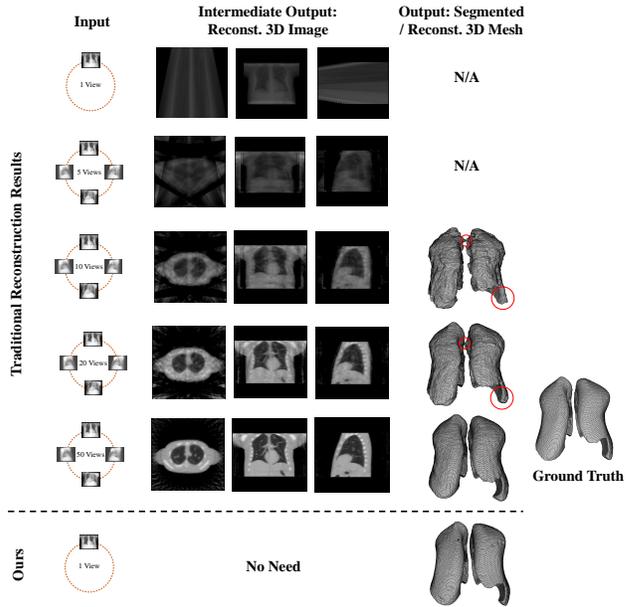}\vspace{-1mm}
		\caption{Qualitative comparison with the traditional SART and our method. The reconstructed 3D-CBCT images by SART from one and five views are unable to be used to segment lungs. SART requires CBCT projections from at least 50 views to reconstruct a good-quality 3D volumetric image such that the corresponding segmented lung model is comparable to our result. Failure parts (e.g., wrong connectivities and not-good shape preserving parts) of SART-based meshes are red-cycled.}\vspace{-8mm}
		\label{fig:tradition}
	\end{center}
\end{figure}

Before deep learning methods applied to 3D reconstruction area, the most common way to acquire 3D organ models from a patient is to first reconstruct 3D-CBCT volumetric image from multiple 2D projections from different views and then segment the organ models from the reconstructed volumetric image. The segmentation quality heavily depends on the number of projections. Very few views severely undermines the reconstructed 3D-CBCT image accuracy, while increasing the views impairs patient health due to higher imaging dose as well as consumes the longer computational time. Our network offers satisfiable 3D lung shape models with only a single-view 3D-CBCT projection. Tab.~\ref{table:tradition} shows that the traditional Simultaneous Algebraic Reconstruction Techniques (SART)~\cite{andersen1984simultaneous} requires at least 10-view projections to reconstruct the 3D lung mesh model and up to 50-view projections to reconstruct the 3D-CBCT volumetric image so as to segment the clean, smooth, and complete 3D lung models to reach the comparable result as ours. Fig. \ref{fig:tradition} shows the qualitative comparison between two methods. It needs to mention that in clinical studies (or during the therapy), it is common to use hundreds of projections to reconstruct a high-quality 3D volumetric image and then process a good-quality 3D organ model.\vspace{-4mm}
\begin{table}[h!]
	\caption{Quantitative comparison between SART (with different numbers of views) and our method.} \vspace{-2mm}
    \label{table:tradition}
	\begin{adjustbox}{width=\columnwidth,center}
\begin{tabular}{|c|c|c|c|c|c|}
	\hline
	Method     & CD     & EMD     & F-score ($\epsilon$ / $1.5\epsilon$)         & IoU    & HD (Mesh) \\ \hline\hline
	1-view  & N/A    & N/A     & N/A             & N/A    & N/A      \\ \hline
	5-view  & N/A    & N/A     & N/A             & N/A    & N/A      \\ \hline
	10-view & 3.3143 & 92.3019 & 0.3999 / 0.6381 & 0.7277 & 0.1888   \\ \hline
	20-view & 2.3458 & 66.1249 & 0.5107 / 0.7470 & 0.9099 & 0.1332   \\ \hline
	50-view & 1.6694 & 43.6351 & 0.6753 / 0.8423 & 0.9433 & 0.0593   \\ \hline
	Ours    & 2.2458 & 48.5677 & 0.4931 / 0.7567 & 0.8880 & 0.0662   \\ \hline
\end{tabular}
\end{adjustbox}\vspace{-4mm}
\end{table}

\subsection{Applications and User Study on Patient Datasets}
To further evaluate the accuracy and usability of the proposed method, our DeepOrganNet has been evaluated in the following studies by some domain experts, including our collaborative radiation oncologists and physicians. The efficiency and accuracy of our method demonstrate its capabilities to explicitly track, reconstruct, and visualize 3D / 4D organ shapes on the fly during the dynamic procedure and therefore it can be employed in the real-time image guided radiation therapy (IGRT).

\subsubsection{3D Lung Shape Reconstruction from Patient Images}
\begin{figure}[t!]
	\begin{center}
		\includegraphics[width=0.9\columnwidth]{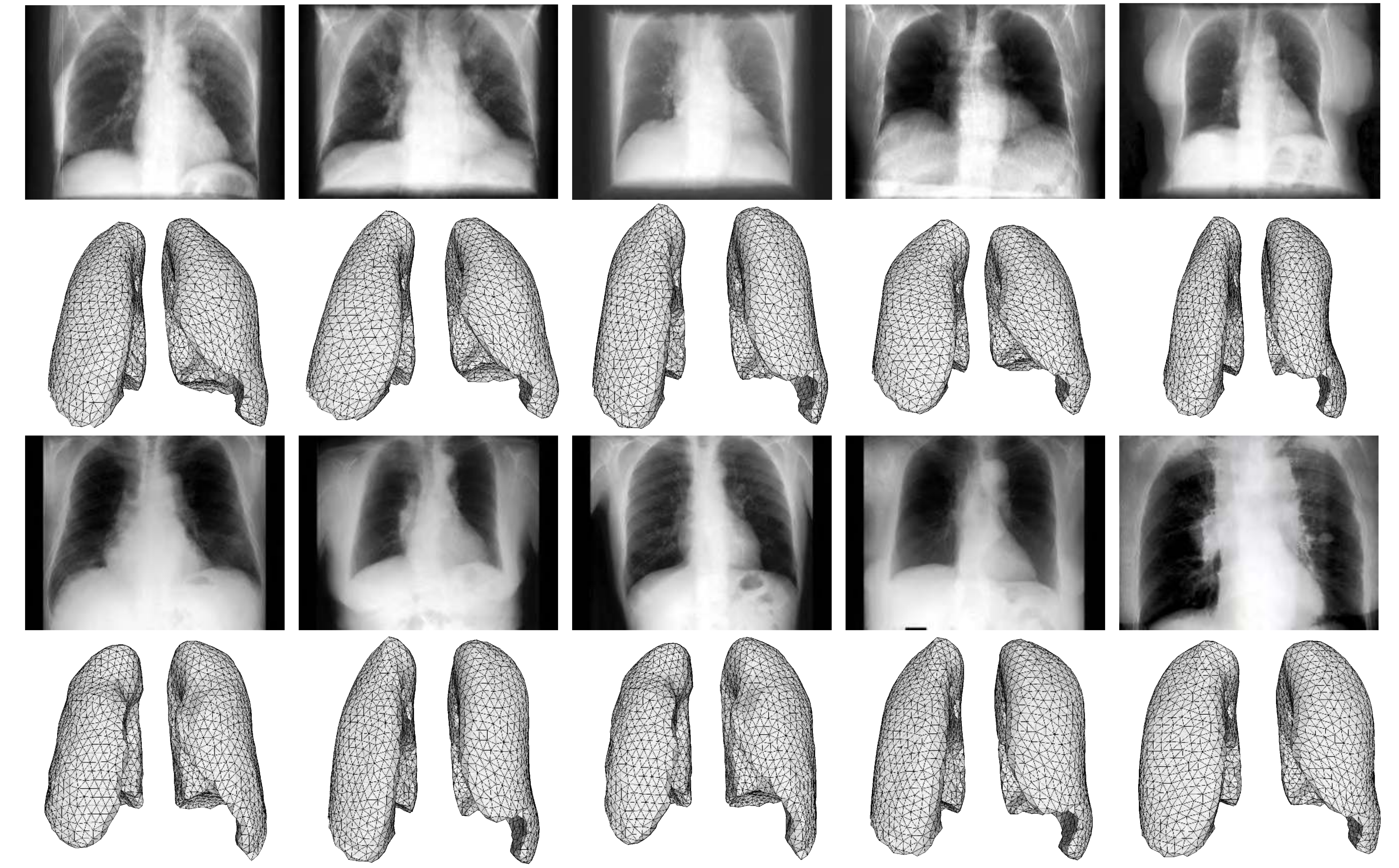}\vspace{-2mm}
		\caption{Top: qualitative visualization results of 3D lung shape reconstruction from single-view phase-0 projections of five cases in DIR-LAB dataset. Bottom: qualitative visualization results of 3D lung shape reconstruction from single-view real X-ray images in JSRT database and real 3D-CBCT patient datasets. These sample results are picked from the challenging cases with large variations of the lung shapes.}\vspace{-8mm}
		\label{fig:real_pick}
	\end{center}
\end{figure}

We first use ten cases of 4D-CTs from DIR-LAB datasets to evaluate the robustness of our method in real applications. For each case, we select the phase-0 of 4D-CTs to compute the front-view CBCT projections using the method in Sec.~\ref{sec:data_generation}. All the generated front-view projections are histogram equalized. We test our network directly on images of all the cases without any fine-tuning. We also further test our model on some single front-view X-ray images from JSRT database~\cite{shiraishi2000development} and some ill-positioned single-view CBCT projections from real lung cancer patient datasets. We can see that our network is capable of describing the shape geometric property and providing a reasonable spatial arrangement in real case even though the images appear to be different from synthetic inputs. Fig.~\ref{fig:real_pick} shows qualitative visualization results of the above datasets, which are examined by domain experts.
\vspace{-2mm}
\subsubsection{4D Lung Shape Reconstruction}
\begin{figure}[t!]
	\begin{center}
		\includegraphics[width=0.75\columnwidth]{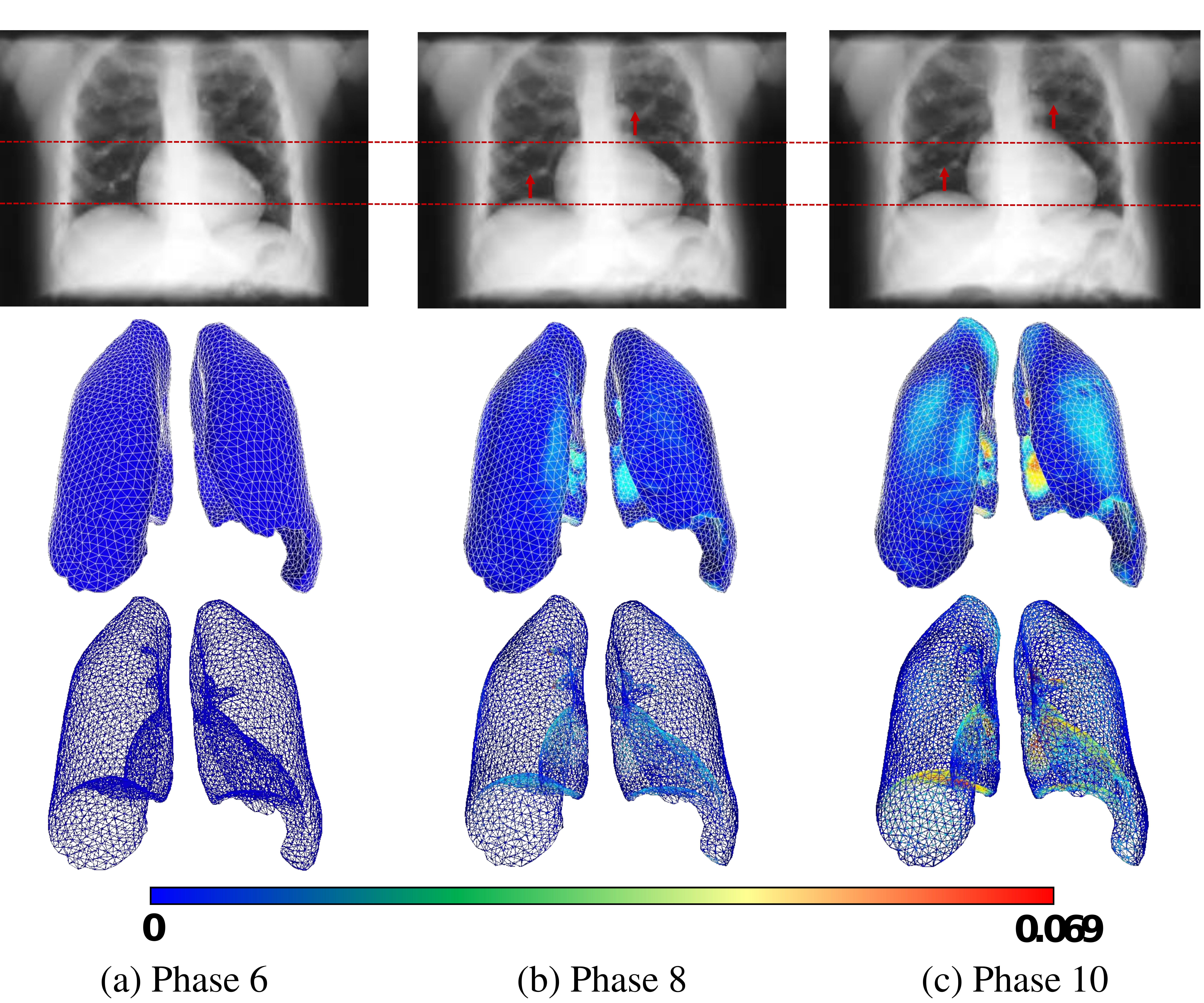}\vspace{-2mm}
		\caption{Three expiration phases of 4D NCAT phantom model. Maximal deformation can be traced according to the red dashed lines across the input 2D images, and the corresponding deformations on the reconstructed 3D mesh models are mapped into a unified colormap range (hotter colors indicate larger deformations). The solid surface and wireframe meshes show the front-view and occluded (diaphragm) deformations, respectively. The deformations of Phases 8 and 10 are computed based on Phase 6 as the reference.}\vspace{-8mm}
		\label{fig:ncat_4d}
	\end{center}
\end{figure}

\begin{figure}[t!]
	\begin{center}
		\includegraphics[width=0.75\columnwidth]{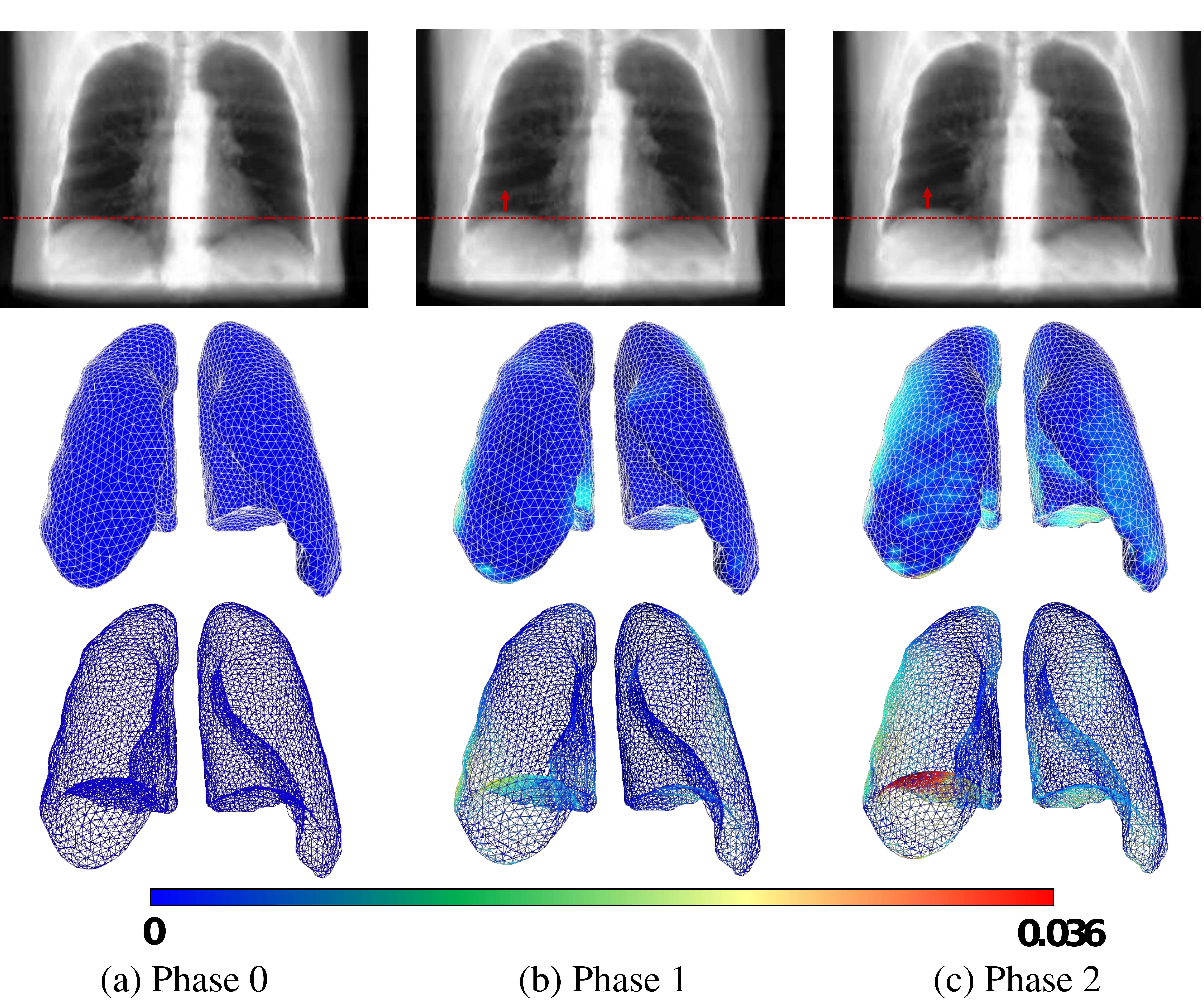}\vspace{-2mm}
		\caption{Three expiration phases of Case 8 in 4D-CT DIR-LAB dataset. Maximal deformation can be traced according to the red dashed line across the input 2D images, and the corresponding deformations on the reconstructed 3D mesh models are mapped into a unified colormap range (hotter colors indicate larger deformations). The solid surface and wireframe meshes show the front-view and occluded (diaphragm) deformations, respectively. The deformations of Phases 1 and 2 are computed based on Phase 0 as the reference.}\vspace{-10mm}
		\label{fig:case8_4d_v2}
	\end{center}
\end{figure}

Instead of inferring the different 3D lung shapes, our network shows potential capability to track and visualize lung shapes along with the dynamic process of breathing. It is extremely important for IGRT procedure to understand the anatomical changes and pinpoint the location of the diseased regions on the fly. By sending a series of 2D front-view 4D-CBCT projections with different phases, our network is capable of capturing the minor changes between phases to describe the breathing tendency and maintaining the shape consistency simultaneously. It is interesting to discover that even some occluded deformations (in the natural images) in the diaphragm areas (bottom part of the lungs) can be extracted and reconstructed from the input single-view X-ray or 4D-CBCT projections. To our knowledge, this is the first time that a single-view reconstruction method can capture that. Fig.~\ref{fig:ncat_4d} and Fig.~\ref{fig:case8_4d_v2} show three expiration phases of a phantom case and a real case in DIR-LAB dataset. The colormaps represent the deformation magnitudes during the breathing. The solid surface meshes and wireframe meshes are used to visualize the front-view and occluded (diaphragm) deformations, correspondingly. Furthermore, the proposed method only takes about 22 milliseconds to generate 4D lung meshes with 5K vertices at each phase, which has a great potential to be used in an on-the-fly targeting system on dynamic scenes in IGRT; however, there is no current method, which can make it on-the-fly.

Since our proposed method outperforms the current methods, an official clinical trial is under arrangement with our collaborative hospital.
\vspace{-2mm}
\section{Conclusion}
In this work, we have proposed the DeepOrganNet, a deep neural network, to generate and visualize high-fidelity 3D / 4D organ shape geometry from single-view medical images in real time. DeepOrganNet has three major components, i.e., feature encoder block, independent deformation block, and spatial arrangement (translation) block. By using the multi-organ template selection and the smooth FFD strategies in the proposed framework, our method can generate high-quality manifold meshing models, which outperforms previous deep learning methods as well as traditional method from the single-view image reconstruction. In medical practice, this work can be used as the key functions for real-time IGRT in order to accurately visualize the patients' organ shapes on the fly, significantly improve the procedure time for patients and doctors, and dramatically reduce the imaging dose during the treatment. Some further interactive techniques based on DeepOrganNet will be developed in collaboration with domain experts.

\textbf{Discussion and Future Work:} In the current framework, the lightweight MobileNets are computational efficient but limit the power of feature extraction in the encoder block. In the future, we will explore some more powerful deep neural networks for the encoder part and collect more 4D lung cancer patient datasets for improving the diversity and scalability of the training and testing for our DeepOrganNet. Although the proposed DeepOrganNet aims to reconstruct multiple 3D organs simultaneously, the current work does only implement for left and right lung organs as an example for justifying the feasibility and extendability of the proposed method. We will extend the framework into more organ reconstructions, such as heart, liver, pancreas, etc., in order to build a real fully DeepOrganNet system at a complicated 3D / 4D scene-level reconstruction. As for 4D scenarios, we have reconstructed each phase independently in the current system, and we will consider to use recurrent neural network and attention-based models for constructing a 4D dynamic organ shape reconstruction deep neural network. It is worth mentioning that the quality of the reconstructed shapes can be further improved by including 2D-view projections from more viewpoints as the input to alleviate shape over- / under-estimation; we will accordingly explore how to balance the computational time (imaging dose) and reconstructed accuracy in the clinical study.
\vspace{-2mm}
\section*{Acknowledgments}
We would like to thank the reviewers for their valuable comments. This work was partially supported by the National Science Foundation under Grant Numbers IIS-1816511, CNS-1647200, OAC-1657364, OAC-1845962, OAC-1910469, the Wayne State University Subaward 4207299A of CNS-1821962, NIH 1R56AG060822-01A1 and ZJNSF LZ16F020002.

\bibliographystyle{abbrv-doi}

\bibliography{template}

\begin{thebibliography}{10}

\bibitem{andersen1984simultaneous}
A.~Andersen and A.~Kak.
\newblock Simultaneous algebraic reconstruction technique {(SART)}: a superior
  implementation of the {ART} algorithm.
\newblock {\em Ultrasonic Imaging}, 6(1):81--94, 1984.

\bibitem{bernardini1999ball}
F.~Bernardini, J.~Mittleman, H.~Rushmeier, C.~Silva, and G.~Taubin.
\newblock The ball-pivoting algorithm for surface reconstruction.
\newblock {\em IEEE Transactions on Visualization and Computer Graphics},
  5(4):349--359, 1999.

\bibitem{botsch2010polygon}
M.~Botsch, L.~Kobbelt, M.~Pauly, P.~Alliez, and B.~L{\'e}vy.
\newblock {\em Polygon mesh processing}.
\newblock AK Peters/CRC Press, 2010.

\bibitem{brock2010reconstruction}
R.~Brock, A.~Docef, and M.~Murphy.
\newblock Reconstruction of a cone-beam {CT} image via forward iterative
  projection matching.
\newblock {\em Medical Physics}, 37(12):6212--6220, 2010.

\bibitem{carreira2016lifting}
J.~Carreira, S.~Vicente, L.~Agapito, and J.~Batista.
\newblock Lifting object detection datasets into {3D}.
\newblock {\em IEEE Transactions on Pattern Analysis and Machine Intelligence},
  38(7):1342--1355, 2016.

\bibitem{Castillo:2009}
R.~Castillo, E.~Castillo, R.~Guerra, V.~Johnson, T.~McPhail, A.~Garg, and
  T.~Guerrero.
\newblock A framework for evaluation of deformable image registration spatial
  accuracy using large landmark point sets.
\newblock {\em Physics in Medicine \& Biology}, 54:1849--1870, 2009.

\bibitem{chang2015shapenet}
A.~Chang, T.~Funkhouser, L.~Guibas, P.~Hanrahan, Q.~Huang, Z.~Li, S.~Savarese,
  M.~Savva, S.~Song, H.~Su, et~al.
\newblock Shape{N}et: an information-rich {3D} model repository.
\newblock {\em arXiv preprint arXiv:1512.03012}, 2015.

\bibitem{chen2008prior}
G.-H. Chen, J.~Tang, and S.~Leng.
\newblock Prior image constrained compressed sensing {(PICCS)}: a method to
  accurately reconstruct dynamic {CT} images from highly undersampled
  projection data sets.
\newblock {\em Medical Physics}, 35(2):660--663, 2008.

\bibitem{choy20163d}
C.~Choy, D.~Xu, J.~Gwak, K.~Chen, and S.~Savarese.
\newblock {3D-R2N2}: A unified approach for single and multi-view {3D} object
  reconstruction.
\newblock In {\em Proceedings of the European Conference on Computer Vision},
  pp. 628--644, 2016.

\bibitem{cignoni1998metro}
P.~Cignoni, C.~Rocchini, and R.~Scopigno.
\newblock Metro: measuring error on simplified surfaces.
\newblock In {\em Computer Graphics Forum}, vol.~17, pp. 167--174, 1998.

\bibitem{deng2009imagenet}
J.~Deng, W.~Dong, R.~Socher, L.-J. Li, K.~Li, and L.~Fei-Fei.
\newblock Image{N}et: A large-scale hierarchical image database.
\newblock In {\em Proceedings of the IEEE Conference on Computer Vision and
  Pattern Recognition}, pp. 248--255, 2009.

\bibitem{ehlke2013fast}
M.~Ehlke, H.~Ramm, H.~Lamecker, H.-C. Hege, and S.~Zachow.
\newblock Fast generation of virtual {X}-ray images for reconstruction of {3D}
  anatomy.
\newblock {\em IEEE Transactions on Visualization and Computer Graphics},
  19(12):2673--2682, 2013.

\bibitem{eigen2014depth}
D.~Eigen, C.~Puhrsch, and R.~Fergus.
\newblock Depth map prediction from a single image using a multi-scale deep
  network.
\newblock In {\em Advances in Neural Information Processing Systems}, pp.
  2366--2374, 2014.

\bibitem{fan2017point}
H.~Fan, H.~Su, and L.~Guibas.
\newblock A point set generation network for {3D} object reconstruction from a
  single image.
\newblock In {\em Proceedings of the IEEE Conference on Computer Vision and
  Pattern Recognition}, pp. 605--613, 2017.

\bibitem{fang2009tetrahedral}
Q.~Fang and D.~Boas.
\newblock Tetrahedral mesh generation from volumetric binary and grayscale
  images.
\newblock In {\em 2009 IEEE International Symposium on Biomedical Imaging: From
  Nano to Macro}, pp. 1142--1145, 2009.

\bibitem{feldkamp1984practical}
L.~Feldkamp, L.~Davis, and J.~Kress.
\newblock Practical cone-beam algorithm.
\newblock {\em Journal of the Optical Society of America A-Optics Image Science
  and Vision}, 1(6):612--619, 1984.

\bibitem{fleute1999nonrigid}
M.~Fleute and S.~Lavall{\'e}e.
\newblock Nonrigid {3-D/2-D} registration of images using statistical models.
\newblock In {\em Proceedings of International Conference on Medical Image
  Computing and Computer-Assisted Intervention}, pp. 138--147, 1999.

\bibitem{fouhey2013data}
D.~Fouhey, A.~Gupta, and M.~Hebert.
\newblock Data-driven {3D} primitives for single image understanding.
\newblock In {\em Proceedings of the IEEE International Conference on Computer
  Vision}, pp. 3392--3399, 2013.

\bibitem{henzler2018single}
P.~Henzler, V.~Rasche, T.~Ropinski, and T.~Ritschel.
\newblock Single-image tomography: {3D} volumes from {2D} cranial {X}-rays.
\newblock In {\em Computer Graphics Forum}, vol.~37, pp. 377--388, 2018.

\bibitem{hoiem2005automatic}
D.~Hoiem, A.~Efros, and M.~Hebert.
\newblock Automatic photo pop-up.
\newblock {\em ACM Transactions on Graphics}, 24(3):577--584, 2005.

\bibitem{howard2017mobilenets}
A.~Howard, M.~Zhu, B.~Chen, D.~Kalenichenko, W.~Wang, T.~Weyand, M.~Andreetto,
  and H.~Adam.
\newblock Mobile{N}ets: Efficient convolutional neural networks for mobile
  vision applications.
\newblock {\em arXiv preprint arXiv:1704.04861}, 2017.

\bibitem{huang2015single}
Q.~Huang, H.~Wang, and V.~Koltun.
\newblock Single-view reconstruction via joint analysis of image and shape
  collections.
\newblock {\em ACM Transactions on Graphics}, 34(4):87, 2015.

\bibitem{islam2006patient}
M.~Islam, T.~Purdie, B.~Norrlinger, H.~Alasti, D.~Moseley, M.~Sharpe,
  J.~Siewerdsen, and D.~Jaffray.
\newblock Patient dose from kilovoltage cone beam computed tomography imaging
  in radiation therapy.
\newblock {\em Medical Physics}, 33(6 Part 1):1573--1582, 2006.

\bibitem{jack2018learning}
D.~Jack, J.~K. Pontes, S.~Sridharan, C.~Fookes, S.~Shirazi, F.~Maire, and
  A.~Eriksson.
\newblock Learning free-form deformations for {3D} object reconstruction.
\newblock In {\em Proceedings of the Asian Conference on Computer Vision},
  2018.

\bibitem{kan2008radiation}
M.~Kan, L.~Leung, W.~Wong, and N.~Lam.
\newblock Radiation dose from cone beam computed tomography for image-guided
  radiation therapy.
\newblock {\em International Journal of Radiation Oncology* Biology* Physics},
  70(1):272--279, 2008.

\bibitem{kar2015category}
A.~Kar, S.~Tulsiani, J.~Carreira, and J.~Malik.
\newblock Category-specific object reconstruction from a single image.
\newblock In {\em Proceedings of the IEEE Conference on Computer Vision and
  Pattern Recognition}, pp. 1966--1974, 2015.

\bibitem{kass1988snakes}
M.~Kass, A.~Witkin, and D.~Terzopoulos.
\newblock Snakes: Active contour models.
\newblock {\em International Journal of Computer Vision}, 1(4):321--331, 1988.

\bibitem{kurenkov2018deformnet}
A.~Kurenkov, J.~Ji, A.~Garg, V.~Mehta, J.~Gwak, C.~Choy, and S.~Savarese.
\newblock Deform{N}et: free-form deformation network for {3D} shape
  reconstruction from a single image.
\newblock In {\em IEEE Winter Conference on Applications of Computer Vision},
  pp. 858--866, 2018.

\bibitem{la2005reduction}
P.~La~Riviere and D.~Billmire.
\newblock Reduction of noise-induced streak artifacts in {X}-ray computed
  tomography through spline-based penalized-likelihood sinogram smoothing.
\newblock {\em IEEE Transactions on Medical Imaging}, 24(1):105--111, 2005.

\bibitem{lamecker2006atlas}
H.~Lamecker, T.~Wenckebach, and H.-C. Hege.
\newblock Atlas-based {3D}-shape reconstruction from {X}-ray images.
\newblock In {\em Proceedings of IEEE International Conference on Pattern
  Recognition}, vol.~1, pp. 371--374, 2006.

\bibitem{li2010real}
R.~Li, X.~Jia, J.~Lewis, X.~Gu, M.~Folkerts, C.~Men, and S.~Jiang.
\newblock Real-time volumetric image reconstruction and {3D} tumor localization
  based on a single x-ray projection image for lung cancer radiotherapy.
\newblock {\em Medical Physics}, 37(6 Part 1):2822--2826, 2010.

\bibitem{li2010single}
R.~Li, X.~Jia, J.~H. Lewis, X.~Gu, M.~Folkerts, C.~Men, and S.~Jiang.
\newblock Single-projection based volumetric image reconstruction and {3D}
  tumor localization in real time for lung cancer radiotherapy.
\newblock In {\em International Conference on Medical Image Computing and
  Computer-Assisted Intervention}, pp. 449--456, 2010.

\bibitem{liu2014wavelet}
X.~Liu, H.~Wang, M.~Xu, S.~Nie, and H.~Lu.
\newblock A wavelet-based single-view reconstruction approach for cone beam
  x-ray luminescence tomography imaging.
\newblock {\em Biomedical Optics Express}, 5(11):3848--3858, 2014.

\bibitem{lorensen1987marching}
W.~Lorensen and H.~Cline.
\newblock Marching cubes: A high resolution {3D} surface construction
  algorithm.
\newblock In {\em ACM SIGGRAPH Computer Graphics}, vol.~21, pp. 163--169, 1987.

\bibitem{pontes2018image2mesh}
J.~Pontes, C.~Kong, S.~Sridharan, S.~Lucey, A.~Eriksson, and C.~Fookes.
\newblock {Image2Mesh}: A learning framework for single image {3D}
  reconstruction.
\newblock In {\em Proceedings of the Asian Conference on Computer Vision},
  2018.

\bibitem{ren2008novel}
L.~Ren, J.~Zhang, D.~Thongphiew, D.~Godfrey, Q.~Wu, S.-M. Zhou, and F.-F. Yin.
\newblock A novel digital tomosynthesis {(DTS)} reconstruction method using a
  deformation field map.
\newblock {\em Medical Physics}, 35(7Part1):3110--3115, 2008.

\bibitem{rubner2000earth}
Y.~Rubner, C.~Tomasi, and L.~Guibas.
\newblock The earth mover's distance as a metric for image retrieval.
\newblock {\em International Journal of Computer Vision}, 40(2):99--121, 2000.

\bibitem{sadowsky2006projected}
O.~Sadowsky, J.~Cohen, and R.~Taylor.
\newblock Projected tetrahedra revisited: A barycentric formulation applied to
  digital radiograph reconstruction using higher-order attenuation functions.
\newblock {\em IEEE Transactions on Visualization and Computer Graphics},
  12(4):461--473, 2006.

\bibitem{saxena2009make3d}
A.~Saxena, M.~Sun, and A.~Ng.
\newblock Make{3D}: Learning {3D} scene structure from a single still image.
\newblock {\em IEEE Transactions on Pattern Analysis and Machine Intelligence},
  31(5):824--840, 2009.

\bibitem{sederberg1986free}
T.~Sederberg and S.~Parry.
\newblock Free-form deformation of solid geometric models.
\newblock {\em ACM SIGGRAPH Computer Graphics}, 20(4):151--160, 1986.

\bibitem{dissertation:Segars2001}
W.~Segars.
\newblock {\em Development and application of the new dynamic {NURBS}-based
  Cardiac-Torso {(NCAT)} phantom}.
\newblock Ph.D. dissertation, University of North Carolina, 2001.

\bibitem{shiraishi2000development}
J.~Shiraishi, S.~Katsuragawa, J.~Ikezoe, T.~Matsumoto, T.~Kobayashi, K.-i.
  Komatsu, M.~Matsui, H.~Fujita, Y.~Kodera, and K.~Doi.
\newblock Development of a digital image database for chest radiographs with
  and without a lung nodule: receiver operating characteristic analysis of
  radiologists' detection of pulmonary nodules.
\newblock {\em American Journal of Roentgenology}, 174(1):71--74, 2000.

\bibitem{siddon1985fast}
R.~Siddon.
\newblock Fast calculation of the exact radiological path for a
  three-dimensional {CT} array.
\newblock {\em Medical Physics}, 12(2):252--255, 1985.

\bibitem{Smith2019}
E.~Smith, S.~Fujimoto, A.~Romero, and D.~Meger.
\newblock {GEOMetrics}: Exploiting geometric structure for graph-encoded
  objects.
\newblock {\em arXiv preprint arXiv:1901.11461}, 2019.

\bibitem{song2007sparseness}
J.~Song, Q.~Liu, G.~Johnson, and C.~Badea.
\newblock Sparseness prior based iterative image reconstruction for
  retrospectively gated cardiac micro-{CT}.
\newblock {\em Medical Physics}, 34(11):4476--4483, 2007.

\bibitem{song2008dose}
W.~Song, S.~Kamath, S.~Ozawa, S.~Alani, A.~Chvetsov, N.~Bhandare, J.~Palta,
  C.~Liu, and J.~Li.
\newblock A dose comparison study between {XVI} and {OBI} {CBCT} systems.
\newblock {\em Medical Physics}, 35(2):480--486, 2008.

\bibitem{szegedy2016rethinking}
C.~Szegedy, V.~Vanhoucke, S.~Ioffe, J.~Shlens, and Z.~Wojna.
\newblock Rethinking the inception architecture for computer vision.
\newblock In {\em Proceedings of the IEEE Conference on Computer Vision and
  Pattern Recognition}, pp. 2818--2826, 2016.

\bibitem{tang20052d}
T.~Tang and R.~Ellis.
\newblock {2D/3D} deformable registration using a hybrid atlas.
\newblock In {\em Proceedings of International Conference on Medical Image
  Computing and Computer-Assisted Intervention}, pp. 223--230, 2005.

\bibitem{wang2006penalized}
J.~Wang, T.~Li, H.~Lu, and Z.~Liang.
\newblock Penalized weighted least-squares approach to sinogram noise reduction
  and image reconstruction for low-dose {X}-ray computed tomography.
\newblock {\em IEEE Transactions on Medical Imaging}, 25(10):1272--1283, 2006.

\bibitem{wang2018pixel2mesh}
N.~Wang, Y.~Zhang, Z.~Li, Y.~Fu, W.~Liu, and Y.-G. Jiang.
\newblock {Pixel2Mesh}: Generating {3D} mesh models from single rgb images.
\newblock In {\em Proceedings of the European Conference on Computer Vision},
  pp. 52--67, 2018.

\bibitem{wu20153d}
Z.~Wu, S.~Song, A.~Khosla, F.~Yu, L.~Zhang, X.~Tang, and J.~Xiao.
\newblock {3D} {ShapeNets}: A deep representation for volumetric shapes.
\newblock In {\em Proceedings of the IEEE Conference on Computer Vision and
  Pattern Recognition}, pp. 1912--1920, 2015.

\bibitem{zhong20163d}
Z.~Zhong, X.~Guo, Y.~Cai, Y.~Yang, J.~Wang, X.~Jia, and W.~Mao.
\newblock {3D-2D} deformable image registration using feature-based nonuniform
  meshes.
\newblock {\em BioMed Research International}, 2016.

\end{thebibliography}
\end{document}